 \documentclass[final,5p,times,twocolumn,authoryear]{elsarticle}

\usepackage{algorithm}
\usepackage{amsmath}
\usepackage{caption}
\usepackage{booktabs}
\usepackage{hyperref}
\usepackage{orcidlink}
\usepackage{mathtools}
\usepackage{xcolor}
\usepackage{tabularx}
\usepackage{diagbox}
\usepackage{listings}
\usepackage{nicefrac}
\usepackage{siunitx}
\usepackage{array,framed}
\usepackage{booktabs}
\usepackage{
  color,
  float,
  epsfig,
  wrapfig,
  graphics,
  graphicx,
  subcaption
}
\usepackage{textcomp,amssymb}
\usepackage{setspace}
\usepackage{latexsym,fancyhdr,url}
\usepackage{enumerate}
\usepackage{algpseudocode}
\usepackage{xparse} 
\usepackage{xspace}
\usepackage{multirow}
\usepackage{csvsimple}
\usepackage{balance}
\usepackage{multicol} 

\usepackage [utf8]{inputenc}

\newcommand{\LONG}[1]
{{}}

\usepackage{
  tikz,
  pgfplots,
  pgfplotstable
}

\usetikzlibrary{
  shapes.geometric,
  arrows,
  external,
  pgfplots.groupplots,
  matrix
}

\pgfplotsset{compat=1.9}

\usepackage{mathtools,}

\DeclareMathAlphabet{\mathcal}{OMS}{cmsy}{m}{n}

\DeclareGraphicsExtensions{%
    .png,.PNG,%
    .pdf,.PDF,%
    .jpg,.mps,.jpeg,.jbig2,.jb2,.JPG,.JPEG,.JBIG2,.JB2}

\begin{document}

\begin{frontmatter}



\title{Protecting Web Applications using a Multimodal AI-based System for Classifying Web Injection Attacks}




\author{Rodrigo Branco, Vinicius Cogo, Ibéria Medeiros \\
fc54457@alunos.ciencias.ulisboa.pt, vvcogo@ciencias.ulisboa.pt, ivmedeiros@ciencias.ulisboa.pt}

\affiliation{organization={LASIGE, Faculty of Sciences of the University of Lisboa},
	addressline={Campo Grande}, 
	city={Lisboa},
	postcode={1749}, 
	state={Lisboa},
	country={Portugal}}


\begin{abstract}
Web applications have been essential to accessing and managing online services. Made available with weaknesses, much due to their rapid development and human error, they are exposed to frequent web injection attacks, leaving security analysts overwhelmed with threats and alerts that often lack critical details for proper risk assessment. Detecting these attacks requires inspecting the payload of network traffic packets, which becomes inaccessible when traffic is encrypted via HTTPS, as encryption hides the content of the traffic. 
%
This paper introduces a Multimodal Artificial Intelligence (AI) system to protect online services that rely on web application security by detecting and classifying web injection attacks, and provides security analysts with information to support attack prevention and incident response.
The system employs unsupervised Machine Learning (ML) to flag potentially anomalous Network Flows (NetFlows).
The corresponding data recorded in Web Server Logs (WebLogs) is then analysed using a heuristic framework based on Natural Language Processing (NLP) and supervised ML to detect and classify web injection attacks.
Finally, the resulting classifications are scrutinised using a Similarity Search Engine (SSE) based on Locality Sensitive Hashing (LSH) to validate them.
The solution handles encrypted traffic by analysing NetFlows and WebLogs, the latter in their natural, unencrypted form, \textcolor{black}{by leveraging NetFlow-level metadata for anomaly detection and analysing web server logs after TLS termination, thereby avoiding any additional in-network traffic decryption}. 
It was evaluated in the classification of eight web injection attack classes. The results demonstrated its ability to accurately identify attacks and their classes, achieving an $f_1$ score of 91\% and an accuracy of 95\%, and offering valuable insights for alert prioritisation.
\end{abstract}

\LONG{
\begin{highlights}
\item A Multimodal AI-based system to detect and classify web injection attacks over NetFlows and WebLogs.
\item Unsupervised Machine Learning (ML) flags potential anomalous HTTP(S) NetFlows.
\item An NLP \& Supervised ML heuristic framework detects and classifies web injection attacks. 
\item A Locality Sensitive Hashing (LSH)-based approach validates attack classification 
\item Gathered attack information helps analysts in attack prevention and incident response.
\end{highlights}
} 


\end{frontmatter}

\section{Introduction}
\label{sec:intro}

Web applications have played a vital role in the evolution of the way we work, enabling easier access and management of (online) services.
The rapid release of their features has often emphasised speed over security, leading to the existence of vulnerabilities and making them attractive targets for attackers seeking unauthorised access to sensitive data (e.g., bank accounts). This rapid release approach increased the number of vulnerabilities by 39\% between 2023 and 2024~\citep{CyberPress:25, cve-date:25}, which aligns with the OWASP~\citep{OWASP21} statement that 94\% of web applications contain injection vulnerabilities, the most prevalent being SQL Injection (SQLi) and Cross-Site Scripting (XSS), and OWASP Top 10 2025~\citep{OWASP10:25} reaffirms it. In 2025, we witnessed a continued growth in the number of vulnerabilities disclosed by 20\%~\citep{cve-date:25}.   

To minimise attackers' actions, the HTTPS protocol has been applied to secure network communication by encrypting network packets, thereby ensuring data confidentiality and thus preventing attackers from interfering with or changing network packets. Although HTTPS protects data in transit, vulnerabilities in endpoint software remain exposed to attackers, so successful attacks can still occur. 
Hence, detecting attackers' actions, such as web application injection attacks, requires inspecting the payload of network traffic packets. Traditional Network Intrusion Detection Systems (NIDS) based on Deep Packet Inspection (DPI)~\citep{DPI-Cheng} are used for such tasks~\citep{shema2010seven}. DPI can, for example, identify SQLi attacks by detecting malicious queries and specific characters~\citep{pramod2015sqli}, but it may cause network slowdowns (i.e., bottlenecks) due to the need to inspect every packet. In addition, it becomes ineffective in HTTPS environments because it requires unencrypted packet content for inspection. To enable DPI in such an environment, proxies are configured to decrypt traffic, but this can lead to security data leakage. 
%

To address network bottlenecks in attack detection, some NIDS solutions rely on Network Flows (NetFlows) or log analysis. 
NetFlows aggregate network-level data (i.e., TCP/IP layer~3) and can operate on HTTPS traffic without requiring decryption. 
Still, they cannot detect application-layer (i.e., packet content of TCP/IP layer~1) attacks such as injections~\citep{Claise2004} and as DPI-based NIDS do~\citep{shema2010seven}. 
Log analysis can help detect anomalies~\citep{nousiainen2009anomaly}, but its effectiveness is limited by large log sizes, often resulting in numerous false positives (FP)~\citep{ber2017}.
Although NIDS can employ Machine Learning (ML) algorithms for anomaly detection~\citep{buczak2016} and improve their effectiveness, these systems struggle with encrypted traffic and lack detailed information about attack types. Such takes that Security Operation Center (SOC) teams often receive too many alerts with insufficient data for vulnerability/attack prioritisation and remediation.
Based on these issues, a desired solution would be the combination of the strengths of various approaches: NetFlow-based anomaly detection for efficiency, supplemented with meaningful attack classification through log analysis with low FP generation, and the provision of useful attack information to support SOC teams in attack prevention and incident response tasks. Furthermore, a solution that works on (un)encrypted traffic, while providing valuable insights into attacks, would enhance the protection of web applications and the services they support. 

In this paper, we present such a solution that leverages Artificial Intelligence (AI) techniques for NetFlows and Web Log (WebLog) analysis in order to detect and classify web injection attacks, such as SQLi and XSS, thereby protecting web applications and the services that rely on them, while keeping SOC teams informed about detected attacks. 
While NetFlows allow the processing of HTTP(S) network traffic and avoid bottlenecks, WebLog analysis enables the inspection of unencrypted content from network traffic \textcolor{black}{after TLS termination and, therefore, avoids the need for any additional in-network traffic decryption}.
To achieve effective detection and classification, we propose a novel Multimodal Artificial Intelligence (AI)-based approach, where $(i)$ unsupervised machine learning is used to detect anomalous NetFlows in HTTP(S) traffic, $(ii)$ followed by log inspection of recorded data caused by the flagged NetFlows using a Heuristic Framework based on Natural Language Processing (NLP) and supervised ML to detect and classify attacks into classes, and $(iii)$ a Similarity Search Engine (SSE) based on Locality Sensitive Hashing (LSH) to validate the attack classification made and reduce false positives and negatives. 
The approach also includes a $(iv)$ recommendation system that leverages information about attacks to prioritise alerts, helping security analysts with tasks such as attack prevention and incident response.

The approach was implemented in a system that supports eight web injection attacks -- SQLi, XSS, Directory Traversal (DT), Template Injection (TI), OS Command Injection (OSCI), CRLF Injection (CRLFI), Code Injection (CI), and XML External Entity (XXE) -- and tested with four web scanners (Wapiti~\citep{wapiti}, Burp~\citep{burp}, ZAP~\citep{zap}, and w3af~\citep{w3af}) to generate malicious traffic. 
The system was evaluated in various scenarios, including known and unknown attacks by the system, as well as benign inputs.
The results demonstrated the system's ability to perform as designed, achieving an $f_1$ score of 91\% and an accuracy of 95\%, and providing accurate and reliable certainty levels for attack classifications, which can help SOC teams perform their actions.

This paper extends our previous work~\citep{branco2024towards} with the following:
(1) a more complete background and related work;
(2) provides more details about the approach, including the challenges and enhanced and new components;
(3) gives more details about the Heuristic Framework (HF) to classify attacks;
(4) presents the SSE LSH-based validator approach for confirming the attacks classified by the HF;
(5) presents the alarm generator approach for helping security analysts in their tasks;
(6) a detailed explanation of the implementation of the system; 
(7) a more complete experimental evaluation that allows for assessing all the system's components.

The main contributions of the paper are:
(1) a novel approach for web application protection that relies on NetFlows and WebLogs analysis to detect and classify web attacks, handling HTTP(S) network traffic and unencrypted WebLogs inspection, and providing meaningful attack information for security threat management. The approach combines NetFlows-based anomaly detection with log-based attack classification and similarity attack validation using AI techniques, including NLP, ML, and LSH;
(2) a set of methods to identify and standardise web injection attacks from recorded payloads into WebLogs for classification and similarity purposes; 
(3) an implementation of the approach in a Multimodal AI-based system and its evaluation for eight web injection attacks, demonstrating its ability to inspect malicious traffic and identify web attacks.

The remaining of the paper is organised as follows: 
Section~\ref{sec:wia} presents the web injection attacks the paper focuses on, Sections~\ref{sec:approach} to~\ref{sec:Att-Class} present the challenges of getting the proposed approach, the approach itself and its design insights. The approach's implementation, instantiation and evaluation are presented in Sections~\ref{sec:impl_inst} and~\ref{sec:evaluation}. 
\textcolor{black}{Section~\ref{sec:real-world} presents the final remarks of the approach's evaluation and its applicability to a realistic deployment.} Section~\ref{sec-background-related} contains the related work, and the paper concludes with Section~\ref{sec:conclusion}.

\section{Web Injection Attacks}
\label{sec:wia}
Web injection attacks try to exploit existing vulnerabilities in web applications by injecting malicious inputs into data fields. Common types include SQLi and XSS, which can lead to unauthorised access, data manipulation, or system compromise~\citep{OWASP-flaws21}. For example, the simple \texttt{' OR 1=1;-- } malicious input of a SQLi attack, when inserted without sanitisation or validation in a query, can cause the undue retrieval of the database's records. On the other hand, for XSS, the injection of a malicious JavaScript (e.g., \texttt{<script>alert("XSS attack")</script>}) and its use, without proper validation, in the output function can allow access to sensitive and private data temporarily stored in the victim's browser (e.g., session cookies).

Next, we briefly present the other six classes we consider in this work. OSCI allows attackers to run commands in a shell of the OS of the web server; DT enables the reading of files from the local file system by exploiting directory traversal; TI lets an attacker exploit template engines by introducing malicious template expressions into the input rendered by engines; CRLFI allows attackers to manage the characters that delimit the termination of a line, which can be used to manipulate a cookie to hijack a session ID, giving attackers unauthorized access to the victim’s session; CI consists of injecting code that is interpreted or executed by the application; and, XXE that allows the manipulation of XML input by attackers.

Web injection attacks are pervasive in network traffic and pose significant risks, passing unnoticed without payloads inspection. Such critical factors make them a focal point in our system’s detection approach.

\LONG{ RW para o fim
\vspace{2mm}
\noindent\textbf{Intrusion Detection and Network Flow.}
\label{sec-netflows-detection}
Anomaly detection is typically approached through \textit{Deep Packet Inspection} (DPI) or \textit{NetFlow-based methods}.  DPI inspects packet content for malicious signatures, identifying specific threats and applications responsible for anomalies~\citep{maghraby2017}. Web application firewalls (WAF) can be seen as a particular DPI for web attacks, as they also inspect payloads~\citep{Gallego:25,Shahid:22}.
Because of this feature, DPI creates a bottleneck on networks with a high volume of traffic and can lead to a denial-of-service~\citep{cisco_vuln}. Moreover, its effectiveness is limited for encrypted traffic, as DPI depends on proxies to decrypt data, which introduces legal and security risks like data exposure~\citep{jarmoc2012ssl}.

On the other hand, NetFlow technology aggregates packets with common properties (e.g., IPs, port numbers) to provide efficient, passive monitoring capabilities~\citep{cisco-netflows,Estan2004,7509765}.
Such aggregation enables NetFlows analysis to surpass the proxy requirement by inspecting only the lower TPC/IP network layers, ensuring high-speed and large-volume networks while mitigating risks associated with DPI. It is suitable for modern encrypted network environments~\citep{Leevy2021}, offering privacy-conscious detection, including botnet command and control (C\&C)~\citep{Subramaniam2021}.
%

\vspace{2mm}
\noindent\textbf{AI for Attack Detection and Classification.}
\label{sec-ml}
Artificial Intelligence (AI) is a broad area that encompasses ML and NLP approaches to achieve its intended goals~\citep{ElNaqa2015}. 
Unsupervised and supervised ML algorithms are the most applied, and we focus on them in this paper. 
Unsupervised ML is used on a large unlabeled dataset to find a structure in unlabelled data and classify it without external validation. This technique is very effective in discovering unseen structures in the data by aggregating similar data into the same group -- cluster --, leaving the unseen data isolated in a cluster. 
LSH is considered a specific method of clustering used for big data tasks. 
Its primary goal is to handle the computational challenges associated with finding the closest data points to a given query point, a task that becomes increasingly difficult as the number of dimensions grows.
LSH enables efficient similarity checks by grouping similar data points, with distance metrics enhancing anomaly detection through clustering~\citep{le2017sonar, Qi2021lsh, bolboac2020}.
On the other hand, supervised ML relies on a dataset with instances labelled into a class/category, then trains a model to understand the pattern behind the labels and predict the class of an unseen sample. Some algorithms used for classification are, for example, Support Vector Machine (SVM)~\citep{svm} and Logistic Regression~\citep{EDGAR201795}. 
NLP, in turn, explores how machines comprehend written and spoken language, including critical application areas, like Text Processing, Syntactic, Semantic and Sentiment Analysis~\citep{Min2023}. NLP is commonly used for text classification, which involves processing the text and capturing its semantics and context to generate a unique vector representation, such as with Doc2Vec. Supervised ML models, for example, can then use this vector for classification problems.

Various works have used AI to detect attacks in network traffic, including inspecting payloads. 
Sarjan \textit{et al.}~\citep{mohanad2021} demonstrated that supervised ML can detect intrusions in inspecting NetFlows. Despite showing promising results, it presents the risk of overfitting, as the labelled data may not adequately represent unseen malicious NetFlows.  
In contrast, unsupervised ML can detect anomalous traffic with NetFlow analysis, as shown by Durão \textit{et al.}~\citep{durao2022discovery}. Their evaluation found that the source IP address differs significantly between benign and malicious traffic.
Das \textit{et al.}~\citep{das2020} proposed NLPIDS that combines NLP with ensemble-based ML to identify anomalous HTTP requests.
LSH has also been applied in anomaly detection, specifically in Industrial 4.0 environments~\citep{Qi2021lsh, bolboac2020}. 
To detect web attacks by payload inspection, 
Seyyar \textit{et al.}~\citep{seyyar2022nlp} uses deep learning, and Kar \textit{et al.}~\citep{kar2016sqligot} applies SVM, both to detect SQLi attacks on HTTP traffic. Irungu \textit{et al.}~\citep{Irungu:23} also detect SQLi attacks on HTTP by measuring metadata of SQL queries (e.g., length) and network traffic (e.g., latency), and feed supervised ML with this data to understand which model performs better.   
Gallego \textit{et al.}~\citep{Gallego:25} also evaluated supervised models over the HTTP traffic inspected by the ModSecurity WAF.  
Vadhil \textit{et al.}~\citep{Vadhil:24} studied which supervised ML model was better suited to the classification of SQLi, XSS and brute force attacks over the CIC-IDS-2017, a well-known and old dataset. The results apparently were good, but were completely biased by the use of an unbalanced dataset containing 98.7\% of benign instances and 1.2\% of those attacks. 
Deep learning models also have been used for network attack detection purposes~\citep{Yi:23}. For example, Zhou \textit{et al.}~\citep{Zhou:25} used deep learning models at 
character-level 
to classify twelve classes of attacks, employing the CSIC 2010 dataset for binary classification and the SR-BH 2020 dataset for multi-class classification.  

Despite these various ML works, none classify the category of an injection attack using NLP and ML, as we propose, i.e., a synergy combination between binary and multi-class models. Regarding LSH, Durão \textit{et al.}~\citep{durao2022discovery} showed that it can be valuable for detecting web application attacks; however, their work only utilised a single LSH database to validate the attack. Our goal is to take this a step further by partitioning the database into multiple entities and devising a classification system for injection attacks.
} 

\section{Challenges \& Approach Overview}
\label{sec:approach}

Our goal is to build a Multimodal AI-based system to detect and classify web injection attacks in HTTP(S) traffic and to provide SOC teams with actionable information about the flagged attacks. Next, we outline the three main challenges to achieving this goal, and then the proposed approach to address them.

\subsection{Challenges}
\label{ssec:challenges} 


\noindent\textit{1) Capturing suspicious HTTP(S) traffic of web attacks:}
The HTTPS protocol encrypts payloads, posing challenges or even invalidating DPI-based inspection since access to unencrypted payloads is required. Furthermore, since DPI systems inspect all packets, they create a bottleneck, which causes delays in detecting and alerting to attacks.
Our idea for dealing with these problems is to use NetFlows to avoid bottlenecks, bypass encryption, and compress multiple packets and provide sufficient metadata for anomaly detection without inspecting the individual contents of the packets. Then, we access WebLogs to retrieve the contents of payloads related to anomalous traffic for later inspection, in which the payloads were registered in web server logs in unencrypted form \textcolor{black}{after TLS termination}.

\noindent\textit{2) Preprocessing payloads for classification:}
Web attacks can be searched in WebLogs using regular expressions~\citep{han2015detection}. However, the studies also show that log attack classification using such an approach has two significant drawbacks on large logs: 1) it tends to create a bottleneck on log search, and 2) it cannot hold the context of the user input, something crucial for the detection of web injection attacks and for avoiding false positives (classifying a benign input as a web attack). 
Our idea to overcome these issues is to retrieve only the logs related to the NetFlows flagged as suspicious, thus significantly reducing the amount of logs to be analysed, and apply a custom attack filter over them, which captures the main features and context of potential attacks.

\noindent\textit{3) Classifying web application attacks:}
Cybersecurity experts have traditionally relied on anomaly detection methods~\citep{gao2017anomaly,seyyar2022nlp} to identify suspicious activity within payloads. However, such methods do not provide the necessary insights for classifying the specific type of attack. Such classification is crucial for SOC teams, as understanding the nature of the attack can minimise the number of alerts they receive and optimise their efficiency.
We propose a comprehensive approach that integrates NLP, ML and LSH techniques to process the preprocessed payloads and classify them into web injection attack classes, minimising false positives and negatives. 
This approach ensures a detailed classification of attacks, thus providing the SOC team with actionable insights to enhance their security posture and incident response capabilities.

\subsection{Overview of the Approach}
\label{ssec:approach} 

\begin{figure}[t]
    \begin{center}
        \includegraphics[scale=0.5]{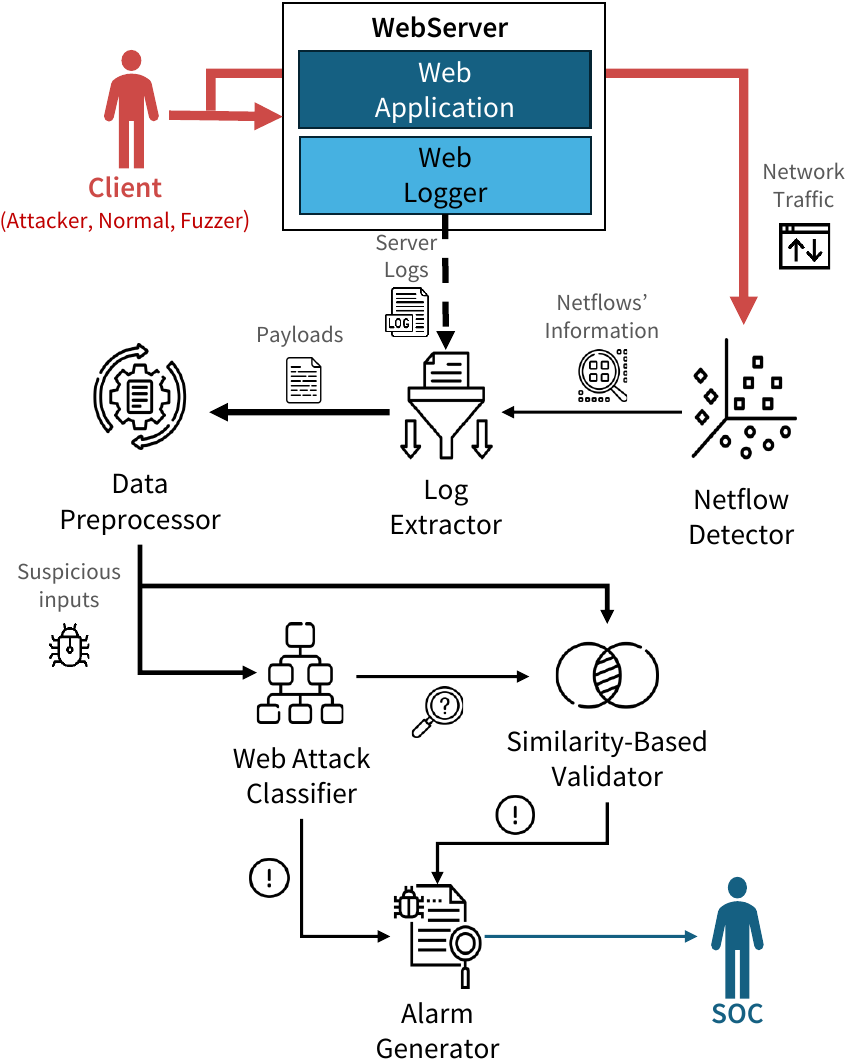}
    \end{center}
    \vspace{-3mm}
    \caption{Architecture of the proposed system. Arrow thickness indicates the decreasing relative data volume throughout the pipeline.}
    \label{fig:approach}
    \vspace{-4mm}
\end{figure}

To pursue our research goal and, therefore, overcome the challenges of Section~\ref{ssec:challenges}, the approach resorts to a combination of ML, NLP and LSH techniques, where unsupervised ML is employed to capture suspicious NetFlows, NLP and supervised ML to detect and classify web attacks on that suspicious traffic, and LSH to detect unseen web attacks that are unknown to the ML classifiers and reduce possible FPs generated by these classifiers. 
Figure~\ref{fig:approach} shows an overview of the approach architecture comprising six modules -- \textit{NetFlow Detector}, \textit{Log Extractor}, \textit{Data Preprocessor}, \textit{Web Attack Classifier}, \textit{Similarity-Based Validator}, and \textit{Alarm Generator} --, divided into the two phases of the approach: \textit{Capturing suspicious traffic} and \textit{Classifying attacks \& Generating alarms}. In the figure, the arrow thickness indicates the amount of data processed from the moment the \textit{Client} initiates an attack until the SOC receives the alarm. This shows that the amount of data decreases throughout the two phases.

The first phase corresponds to the first three modules, which inspect network traffic represented as NetFlows to capture abnormal traffic, gather the WebLogs associated with it, and prepare the payloads they contain for further processing by the next phase.
In the second phase, which corresponds to the other three modules, the already pre-processed payloads enter in an NLP, ML, and LSH pipeline to identify web injection attacks they may carry. If any attack is identified, alarms are generated to the SOC.
%
Sections~\ref{sec:Capt-Traff} and \ref{sec:Att-Class} detail the phases, and Section~\ref{sec:impl_inst} their implementation into a Multimodal AI-based system for web injection attack classification. 
\textcolor{black}{Section~\ref{sec-background-related} discusses the most important and related works and states in which our proposal improves the web attack detection and classification, and innovates compared with other solutions.}

\section{Capturing Suspicious Traffic}
\label{sec:Capt-Traff}

\subsection{NetFlow Detector} 

This module aims to detect suspicious anomalous HTTP(S) network traffic associated with malicious activity (e.g., web scanners). To do so, it captures the traffic generated by the Client-WebApp in the form of NetFlows, i.e., it captures the PCAPs of the bi-directional traffic that occurs over a period of time (e.g., 3 minutes) and converts them into a NetFlow. NetFlows have a set of features provided by the Network and Transport layers of TCP/IP protocol, such as source and destination IP and ports, as well as calculated features, such as the number of packets and bytes exchanged during that period. These features are converted to numbers for the ML algorithms to use. 
Afterwards, NetFlows are clustered using an unsupervised ML algorithm, such as k-Means or DBSCAN, yielding distinct clusters with similar characteristics. 
Clusters containing scarce NetFlows are deemed suspicious (i.e., anomalous and likely an attack) because
\textcolor{black}{the majority of NetFlows representing benign behaviour in the web application are aggregated into a single large cluster~\citep{Sacramento:18,Sacramento:20}}. These smaller clusters are marked for processing by the \textit{Log Extractor} module to retrieve the associated logs from the Web Logger on the WebServer entity.
At this point, the approach has significantly reduced the data for further analysis, leaving only a few NetFlows to be inspected and avoiding the bottleneck. \textcolor{black}{These NetFlows, at this point, are considered as being suspicious attacks and so flagged as candidates for further inspection.}
%

\subsection{Log Extractor} 

Web injection attacks are strictly related to the injection of user inputs into GET and POST requests. 
The web server receives these requests and logs them. 
In the case of HTTPS, the requests are received \textcolor{black}{in cleartext after TLS termination, i.e., at the end of the HTTPS channel}.
We are particularly interested in obtaining these requests in the event that suspicious NetFlows contain some of them. The \textit{Log Extractor} module has this role, and, therefore, it filters the Web Logger's logs to retrieve the data sent by the Client to the WebApp regarding the marked suspicious NetFlows. The logs, in addition to containing user input data, also contain the URL and parameters involved in the suspected attack, such as the input fields of the web page, which are important to the approach to understanding the context of the attack on the target web application.

To extract the corresponding NetFlows' logs, the module matches the \texttt{$<$datetime, source IP, destination IP$>$} tuples of NetFlows with the logs, considering a \texttt{datetime} deviation between the time when traffic is captured and the data is logged. This deviation is managed through a small threshold (e.g., 20 seconds) to ensure that all the logs related to the identified anomalous traffic are captured. Additionally, the module discards logs that do not include a payload, i.e., data that does not come from GET and POST requests. Lastly, as this data comes in chunks, it rebuilds the payloads by identifying the GET and POST requests and joining all the parts. 

\subsection{Data Preprocessor}
\label{sec:data-prepro-detail} 

This module prepares the suspicious payloads by capturing essential injection attack patterns for an accurate identification and classification of attacks.
It produces two standardised payload outputs, tailored for use by the \textit{Web Attack Classifier} and \textit{Similarity-Based Validator} modules, once they operate with different AI strategies. To achieve this, it employs a three-step pipeline, explained next and accompanied by the showcase illustrated in Listing~\ref{lst:preprocessing-example} for the payload \textbf{Pay}.  
\vspace{2mm}
\noindent\textit{\textbf{S1:} User input extraction:} 
extracts user input values from payloads, disregarding URL-specific features (e.g., parameters or input fields). However, each processed payload is associated with its corresponding URL, which is stored in a database and used to generate an alarm if needed.
This input isolation allows for a more adaptable approach to any web application. Note that the injected inputs from web scanners are provided from a database they contain, and are applied in any input field of the web applications they are attacking, regardless of the web application context. This means that if a web scanner explores two different applications, it uses the same inputs.   

\noindent\textit{\textbf{S2:} Dummy values removal:} 
filters out dummy values, i.e., strings lacking meta-characters typical of attacks. Dummy values are detectable because the Web Logger encodes all metacharacters in HTML URL format (e.g., a \textit{slash} (``{\small\texttt{/}}'') is encoded into {\small \texttt{\%2F}}). In the listing, for example, the strings {\small\texttt{hi}}, {\small\texttt{password}}, and {\small\texttt{submit}} are dummy values, and so removed.

\noindent\textit{\textbf{S3:} Standardization:} 
standardises suspicious values using a \textit{Custom Dictionary} that maps common attack patterns (e.g., SQL commands, JavaScript, Python, PHP code, Directory Traversal substrings) to specific tokens; e.g., the SQL command \texttt{and} is tokenised as {\small \texttt{AND}}. Numeric values are replaced with the token {\small\texttt{NUM}} (e.g., in the listing, {\small\texttt{158}} is replaced by {\small\texttt{NUM}}), and unrecognised values are replaced with a general token {\small \texttt{TOKEN}}. To ensure each token corresponds uniquely to one value, each recognised value is represented in uppercase form (e.g., \texttt{len} becomes {\small \texttt{LEN}}). Exceptions include file extensions (e.g., \textit{.txt} and \textit{.php}), which are represented by {\small \texttt{FILE}} to indicate file use without specifying the exact file.

The result of \textbf{S3} in Listing~\ref{lst:preprocessing-example} is the one for \textit{Web Attack Classifier}. For \textit{Similarity-Based Validator}, the output is similar but without the spaces and with uniform token sizes, due to the similarity-based algorithm’s constraints (see Section~\ref{sec:implementation}). 

\begin{lstlisting}[mathescape, breaklines, numbers=none, caption=Example of the preprocessing of a payload (Pay over S1 to S3) and its vector representation (Vec) generated by the Doc2Vec NLP algorithm., label={lst:preprocessing-example}, escapeinside={(*@}{@*)}]   
(*@\textbf{Pay:}@*) login=hi&password=password&num_secure=158%22%29%20and%2082%3d82%20and%20%28%2216%22%3d%2216&submit=submit
                                $\downarrow$
(*@\textbf{S1:}@*)  ["hi", "password", "158%22%29%20(*@\textcolor{red}{and}@*)%2082%3d82%20(*@\textcolor{red}{and}@*)%20%28%2216%22%3d%2216", "submit"]
                                $\downarrow$
(*@\textbf{S2:}@*)  ["158%22%29%20(*@\textcolor{red}{and}@*)%2082%3d82%20(*@\textcolor{red}{and}@*)%20%28%2216%22%3d%2216"]
                                $\downarrow$
(*@\textbf{S3:}@*)  NUM (*@\textcolor{blue}{\%22}@*) (*@\textcolor{blue}{\%29}@*) (*@\textcolor{blue}{\%20}@*) (*@\textcolor{red}{AND}@*) (*@\textcolor{blue}{\%20}@*) NUM (*@\textcolor{blue}{\%3d}@*) NUM (*@\textcolor{blue}{\%20}@*) (*@\textcolor{red}{AND}@*) (*@\textcolor{blue}{\%20}@*) (*@\textcolor{blue}{\%28}@*) (*@\textcolor{blue}{\%22}@*) NUM (*@\textcolor{blue}{\%22}@*) (*@\textcolor{blue}{\%3d}@*) (*@\textcolor{blue}{\%22}@*) NUM
                                $\downarrow$
(*@\textbf{Vec:}@*) [0.0399 0.1332 -0.0544 ... 0.0021 0.9311 -0.1202 -0.2332]
\end{lstlisting}
\vspace{-3mm}

\section{Classifying attacks \& Generating alarms}
\label{sec:Att-Class}

This phase is the core of our proposed approach, as it is responsible for processing the already standardised suspicious user inputs to $(i)$ detect attacks and classify them into a web injection attack class while also outputting the certainty level of its classification, $(ii)$ validate the classification to reduce false positives and negatives, and $(iii)$ generate an alarm if needed.
The classification's information and certainty will then be transmitted to the SOC teams (in the form of alarms), giving them confidence in the security decisions they make.

To conduct the attack classification and validation, the phase relies on a \textit{Heuristic Framework} (HF) that integrates NLP, supervised ML, and LSH algorithms, along with a set of custom heuristics. 
In the end, the framework categorises suspicious user inputs into one of the following four states, accompanied by an alarm sent to SOC, containing all the information obtained during the process.

\begin{itemize} 
    \item \textbf{$\mathit{ATT}$}: The input is considered an \texttt{ATTack}, and the attack class is predicted with high certainty. 
    \item \textbf{$\mathit{UNC}$}: The instance is considered an attack and belongs to one of the known classes, but it can not be precisely determined, i.e., the class is given with uncertainty (\texttt{UNCertain-class}). 
    \item \textbf{$\mathit{AON}$}: 
    The heuristic can not determine with a reasonable level of certainty whether the instance is an attack of a class it knows (\texttt{Attack-Or-Not}). 
    \item \textbf{$\mathit{NAT}$}: The instance is considered to belong to no known attack class (\texttt{Non-ATtack}). This does not guarantee the instance is benign; it merely indicates that the system does not classify it as a known attack.
\end{itemize}

Next, we present the heuristic framework, alongside the modules that integrate this phase.

\subsection{Web Attack Classifier}
\label{subsec:web-att-det} 

The Web Attack module performs the attack detection and classification by processing the user inputs with NLP and supervised binary- and multi-class ML algorithms. 
While NLP aims to represent suspicious user inputs as numeric vectors, such as the one shown in the \textbf{Vec} line of Listing~\ref{lst:preprocessing-example} and in a format that ML algorithms can process, ML algorithms classify the vectors into an attack class. Then, heuristics determine the final attack class among those obtained from ML and provide the certainty level of the classification.

The reasoning behind the binary- and multi-class ML algorithms within the framework is to achieve a more realistic and trustworthy classification. 
We aim to achieve a consensus between different classifiers, resulting from several classifications of the same instance, rather than relying on a single classifier. 
Thus, within the framework, their classifications are compared and combined to justify the final classification. 
\textcolor{black}{The proposed consensus heuristic can be seen as a rule-based classifier fusion mechanism~\cite{kuncheva2004combining}. Unlike conventional bagging, boosting, and stacking ensembles, it does not aggregate bootstrap-trained models, train models sequentially, or use a learned meta-classifier. Instead, binary and multiclass probability outputs are fused through explicit rules based on probability thresholds, class agreement, and uncertainty states.}

The idea relies on having a binary classifier for each attack class supported by the framework, as well as a multi-class classifier for each group of four attack classes. 
Thus, we ensure that binary- and multi-class classifiers can give, respectively, a minimum certainty level of 0.5 and 0.25 probabilities in their classifications. Our choice of having a multi-class classifier of a maximum of four attack classes relies on this minimum level because we cannot ensure a reliable certainty level with probabilities less than 0.25 (e.g., a multi-class classifier for six or eight classes).
For example, if the framework is configured to identify four attack classes (e.g., DT, SQLi, CI, and XXE), it will integrate four binary classifiers, one for each class that each of them will classify the input vector to the class it represents, and one multi-class classifier that will classify the input vector for all four classes. 
As the number of classifiers varies as the number of attack classes supported, it is easy to determine the number of binary classifiers, while the number of multi-class classifiers depends on multiples of four (maximum and if possible) or, otherwise, of three. For instance, in the case of eight classes, the framework will integrate this number of binary classifiers and two multi-class classifiers. 

\subsubsection{Heuristic for Classification} 
\label{subsubsec:heu-detail} 

The multi-classification models return a probability vector $v_m$, indicating the likelihood of an instance belonging to each class. In contrast, binary models return two probabilities: positive (belonging to the class) and negative (not belonging to the class). To create the binary vector for the heuristic, we gather the positive probabilities from all binary models and form $v_b$, which aligns with $v_{m}$ classes.

The heuristic processes these vectors to classify the suspicious user inputs into one of the states stated above. It performs a three-step pipeline: first, assigning a temporary state to the instance, second, classifying it by attack class, and third, finding a consensus between instance classifications. These steps ensure the system efficiently handles uncertainties and minimises incorrect classifications that may arise from malicious inputs not matching any known class. Thereby, they help reduce false positives and negatives.

\noindent\textbf{Step 1: Temporary State of a Probability Vector.} 
\label{def:multi-tuple}
Indicates whether the vector can be clearly assigned to a class, and therefore has a defined temporary state.  
Thus, given a vector $v$ composed of probabilities $p_{c_i}$ of an instance belonging to each class $c_i$, and the threshold $t$ that defines the certainty of the vector to be assigned to class $c_i$, the temporary state of a probability vector, $ts(v)$, is determined by Equation~\ref{eq:ts}.
\vspace{-2mm}

{\small
\begin{equation}\label{eq:ts}
    ts(v)=
    \begin{cases}
        \mathit{DEF}, & \max(v) > t \\
        \mathit{UND}, & \max(v) \leq t                 
    \end{cases}
\end{equation}
}

\noindent
where: \\
    $\bullet$ $v =(p_{c_1},\ldots,p_{c_n})$ a vector of $n$ $p_{c_i}$ probabilities, each one corresponding to a distinct class $c_i$ out of $n$ classes, such as $1 \leq n < \infty$. \\  
    $\bullet$ $t$ is a threshold, such as $\frac{1}{n} < t \leq 1$.

The minimum threshold $t$ allowed is 0.25 for multi-class vectors $v_m$, while it is 0.5 for binary vectors $v_b$. This means that two thresholds are defined, one for multi-class vectors ($t_1$) and another for binary vectors ($t_2$).

\noindent\textbf{Step 2: Class of a Probability Vector.}
\label{def:gen-class}
Given a probability vector $v$, there is a function $cls$, that represents the class corresponding to a probability $p_{c_i}$, such as $cls(p_{c_i}) = c_i$.

\begin{figure}[t]
\begin{center}
\includegraphics[scale=0.55]{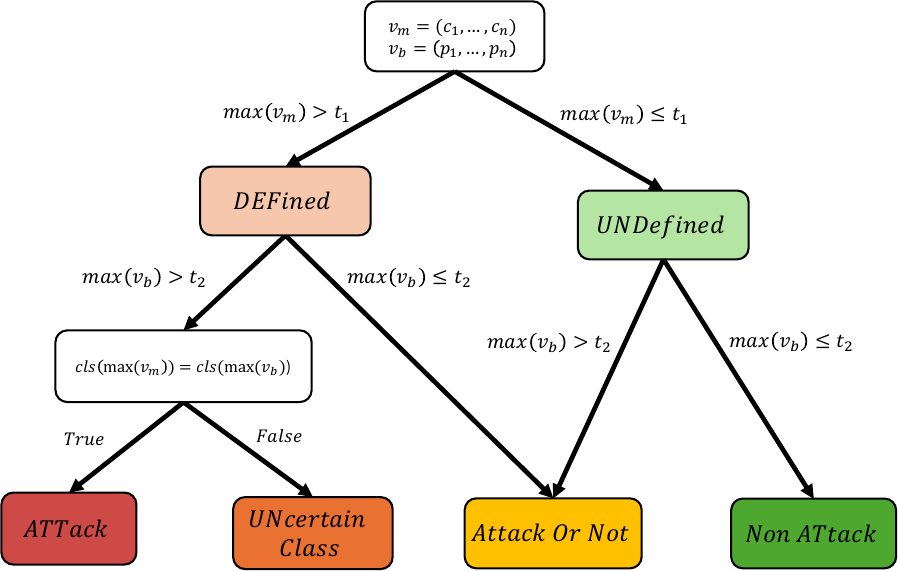}
\end{center}
\vspace{-3mm}
\caption{Illustrated Heuristic detailed in Equation~\ref{eq:states}}
\label{fig:heuristic}
\vspace{-5mm}
\end{figure}

\noindent\textbf{Step 3: Consensus between Temporary States.}
After creating the temporary states and getting the classes of the multi- and binary classification probability vectors, a consensus is performed to determine the final State using Equation~\ref{eq:states}, which details the final heuristic branch state $s$ of the probability vectors $v_m$ and $v_b$, i.e., $s(v_m, v_b)$.
Figure~\ref{fig:heuristic} illustrates a representation of Equation~\ref{eq:states}. If both probability vectors are considered $DEF$ and the class of the maximum of both probability vectors is the same, then we consider the instance an $\mathit{ATT}$. If they are both $DEF$ but the class is not the same, it is an $\mathit{UNC}$. If only one of the probability vectors is $DEF$, then it is an $\mathit{AON}$. If both are $\mathit{UND}$, then the final state is $\mathit{NAT}$.

{\small
\begin{equation}\label{eq:states}
	s(v_m, v_b)
	= 
	\begin{dcases}
		\mathit{ATT}, & (ts(v_m)=DEF \wedge ts(v_b)=DEF) \,\ \wedge \\ 
		     & cls(\max(v_m)) = cls(\max(v_b))  \\
		\mathit{UNC}, & (ts(v_m)=DEF \wedge ts(v_b)=DEF) \,\ \wedge \\ 
		     & cls(\max(v_m)) \neq cls(\max(v_b)) \\
		\mathit{AON}, & (ts(v_m)=DEF \wedge ts(v_b)=UND) \,\ \vee \\ 
		     & (ts(v_m)=UND \wedge ts(v_b)=DEF)                   \\
		\mathit{NAT}, & ts(v_m)=UND \wedge ts(v_b)=UND        
	\end{dcases}
\end{equation}
}


An example to illustrate the application of heuristic is presented next for the instance of \textbf{Vec} line of Listing~\ref{lst:preprocessing-example}, and supposing $v_{m} = (0.10, 0.50, 0.15, 0.25)$ and $v_{b} = (0.23, 0.95, 0.32, 0.47)$ probability vectors for the following four attack classes (DT, SQLi, CI, XXE), and $t1 = 0.35$ and $t2 = 0.50$ thresholds. 

\noindent
    $\bullet$ $max(v_{m}) = 0.50 > t_1 \implies ts(v_{m}) = DEF$ \\
    $\bullet$ $max(v_{b}) = 0.95 > t_2 \implies ts(v_{b2}) = DEF$ \\
    $\bullet$ $cls(max(v_{m})) = cls(max(v_{b}))$; so the instance is considered an attack ($ATT$) and its class is an SQLi.

\subsubsection{Heuristic for Multiple-Vector Classification} 
\label{subsubsec:heu-detail} 

In the case more than four attack classes is supported by the framework, and so various multi-class classifiers it integrates, the Equation~\ref{eq:states} is applied to each $(v_{m_i}, v_{b_i})$ pair of the same classes, resulting in a set $S=(s_1,...,s_m)$ that contains the consensus of temporary states of the pairs, with $s_i=s(v_{m_i}, v_{b_i})$. The final State for the instance in evaluation is then achieved by the consensus of $s_i$ belonging to $S$ through Equation~\ref{eq:states_S}, $hs(s_1,...,s_m)$. 
According to the equation, if only one heuristic branch identifies the state as $\mathit{ATT}$, the final state is set to $\mathit{ATT}$. If at least two branches classify it as $\mathit{ATT}$ or one branch identifies it as $\mathit{UNC}$, the state becomes $\mathit{UNC}$. This rule structure creates a cascade effect, with $\mathit{ATT}$ prioritised over $\mathit{UNC}$, which is then prioritised over $\mathit{AON}$, and finally $\mathit{AON}$ over $\mathit{NAT}$.


{\footnotesize
\begin{equation}\label{eq:states_S}
    \begin{cases}
    \!
    \begin{split}
        \mathit{ATT},  \,\ \exists s_i \in S, \forall s_j \in S \setminus \{s_i\}: &  \,\ s_i=ATT \wedge s_j \neq ATT \\
        \mathit{UNC},  \,\ \exists s_i,s_j \in S, \forall s_k \in S \setminus \{s_i\}: & \,\ (s_i=ATT \wedge s_j=ATT) \,\ \vee \\
        & \,\ (s_i=UNC \wedge s_k \neq UNC) \\
        \mathit{AON},  \,\ \exists s_i,s_j \in S, \forall s_k \in S \setminus \{s_i\}: & \,\ (s_i=UNC \wedge s_j=UNC) \,\ \vee \\
        & \,\ (s_i=AON \wedge s_k \neq AON) \\
        \mathit{NAT},  \,\ \exists s_i,s_j \in S, \forall s_k \in S \setminus \{s_i\}: & \,\ (s_i=AON \wedge s_j=AON) \,\ \vee \\
        & \,\ (s_i=NAT \wedge s_k \neq NAT) 
    \end{split}
    \end{cases}
\end{equation}
}

A concrete example of this scenario and equation is provided by Equation~\ref{eq:rule}, which utilises two multi-class classifiers. In this scenario, the heuristic of Equation~\ref{eq:states} is applied twice, one for each pair, resulting $s_1=s(v_{m1}, v_{b1})$ and $s_2=s(v_{m2}, v_{b2})$. 
Then, Equation~\ref{eq:rule} is applied to $s_1$ and $s_2$, resulting $hs(s_1, s_2)$. In the equation, for example, $ATT_1$ is the result $s_1$, thus indicating that the heuristic branch of $s(v_{m1}, v_{b1})$ identifies the instance as an attack, $ATT$, and thereby represented as $ATT_1$ for that pair. 

{\small
\begin{equation}\label{eq:rule}
    hs(s_1, s_2)
    = 
	\begin{dcases}
		\mathit{ATT}, & \text{if } s_1=\mathit{ATT}_1 \wedge s_2= \neg \mathit{ATT}_2 \\
		\mathit{UNC}, & \text{if } (s_1=\mathit{UNC}_1 \wedge s_2=\neg \mathit{UNC}_2) \\
             & \vee (s_1=\mathit{ATT}_1 \wedge s_2=\mathit{ATT}_2) \\
		\mathit{AON}, & \text{if } (s_1=\mathit{AON}_1 \wedge  s_2=\neg \mathit{AON}_2) \\
             & \vee (s_1=\mathit{UNC}_1 \wedge s_2=\mathit{UNC}_2) \\
		\mathit{NAT}, & \text{if } (s_1=\mathit{NAT}_1 \wedge  s_2=\neg \mathit{NAT}_2) \\
             & \vee (s_1=\mathit{AON}_1 \wedge s_2=\mathit{AON}_2) 
	\end{dcases}
\end{equation}
}


\subsubsection{Getting the Final Class and Certainty Level} 
\label{subsubsec:heu-detail} 

Every instance under evaluation must be associated with a class, even if it is not identified as an attack, including those classified as $\mathit{NAT}$.
When an instance is classified as an attack ($\mathit{ATT}$), the system directly assigns it a class, $c_p$, based on the resulting prediction, which then indicates the level of certainty that the system affirms the instance is an attack. This class corresponds to one of the $n$ attack classes supported by the framework. In contrast, the states $\mathit{UNC}$ and $\mathit{AON}$ do not directly predict a class but rather indicate a class tendency, denoted as $c_t$, with $\mathit{UNC}$ indicating a stronger tendency than $\mathit{AON}$. Lastly, the state $\mathit{NAT}$ also may indicate a class tendency $c_t$, even though it may be weak.
  
The $\mathit{UNC}$ state occurs when $v_m$ and $v_b$ both suggest distinct classes. Thus, to determine $c_t$, $v_m$ and $v_b$ must be normalised as they represent different models with different thresholds ($t_1=0.25$ for multi-class models and $t_2=0.5$ for binary models). To do so, the maximum probabilities of $v_m$ and $v_b$ are normalised to identify the more probable class (Equation~\ref{eq:normalisation}).
For example, if $max(v_m) = 0.72$ and $max(v_b) = 0.80$, normalization results in $0.62$ and $0.6$, respectively, so $c_t = cls(max(v_m))$, the more certain class, and $max(v_m) = 0.72$, the certainty level.

{\small
\begin{equation}\label{eq:normalisation}
	X_{new} = \frac{X - X_{min}}{X_{max} - X_{min}}
\end{equation}
}

The $\mathit{AON}$ state occurs when only one of $v_m$ or $v_b$ meets its threshold (i.e., $t_1$ or $t_2$), so $c_t$ is based on the vector that meets the threshold. The $\mathit{NAT}$ state applies when neither vector exceeds its threshold, in which case normalisation is applied, as in Equation~\ref{eq:normalisation}, to determine the best class.

\subsection{Similarity-Based Validator}
\label{subsec:similarity-based-validator}

While the Web Attack Classifier module predicts potential threats, it is susceptible to false positives and negatives due to the complexity of injection attacks. Given the infinite variations in attack execution, a second filter module is essential to confirm the classifier’s accuracy. The \textit{Similarity-Based Validator} was designed to address this need.
This module employs LSH, an unsupervised similarity metric, to reduce system bias. After the Web Attack Classifier identifies a potential attack, the module compares it against known attacks stored in $n$ modular LSH databases, each corresponding to a specific attack class. 
To perform its actions, the module receives the instance under evaluation in the format of the LSH string generated by the Data Preprocessor (see Section~\ref{sec:data-prepro-detail}), queries the LSH databases, and classifies the instance into one of the attack classes, denoted as $c_{LSH}$. As detailed earlier, instances classified with the state $\mathit{ATT}$ are associated with the class $c_p$, while $\mathit{UNC}$, $\mathit{AON}$, and $\mathit{NAT}$ states are linked to class $c_t$. The module follows the rules in Equation~\ref{eq:lsh-rule} to assign the Similarity-Based Validator state $ss(hs,c_p,c_t,c_{LSH})$, following the logic from Equation~\ref{eq:states_S}.

If an instance is classified as $\mathit{ATT}$ and $c_p$ matches $c_{LSH}$, the attack is confirmed. Conflicting results between the Web Attack Classifier and the Similarity-Based Validator result in a classification of $\mathit{UNC}$. If the instance is classified as $\mathit{UNC}$ or $\mathit{AON}$, and $c_t$ matches $c_{LSH}$, the module confirms the attack with the state $\mathit{ATT}$. Otherwise, the state remains unchanged. Both modules hold equal weight in the assessment.
The state $\mathit{NAT}$ cannot change to $\mathit{ATT}$ even if both modules agree. Instead, it is reassigned to $\mathit{AON}$, i.e., if both modules agree on $\mathit{NAT}$, indicating the instance does not belong to any known class, the final state will be $\mathit{AON}$.

{\small
\begin{equation}\label{eq:lsh-rule}
    {\rotatebox[origin=c]{90}{$ss(hs,c_p,c_t,c_{LSH})$}}
    =
	\begin{dcases}
		\mathit{ATT}, & (hs=\mathit{ATT} \wedge c_p = c_{LSH}) \,\ \vee \\
        & ((hs=\mathit{UNC} \vee hs=\mathit{AON}) \wedge c_t = c_{LSH}) \\
		\mathit{UNC}, & (hs=\mathit{ATT} \wedge c_p \neq c_{LSH}) \,\ \vee \\ 
        & (hs=\mathit{UNC} \wedge c_t \neq c_{LSH})  \\
		\mathit{AON}, & (hs=\mathit{AON} \wedge c_t \neq c_{LSH}) \,\ \vee \\ 
        & (hs=\mathit{NAT} \wedge c_t \neq c_{LSH})\\
		\mathit{NAT}, & hs=\mathit{NAT} \wedge c_t \neq c_{LSH} \\ 
	\end{dcases}
\end{equation}
}


\subsection{Alarm Generator}
\label{subsec:alarm-generator-detail}

The Alarm Generator module creates alerts based on the CWE vulnerability catalog~\citep{cwe}, which assigns each vulnerability a unique identifier. Using the state and class of the detected attack, the system retrieves the relevant CWE identifier to assess the technical impact, covering availability, integrity, confidentiality, and likelihood. If the detected attack has a high-impact CWE, the module triggers a high-priority alarm to the SOC team.

While the $\mathit{NAT}$ state is generally assigned a low classification level, the priority may vary depending on the vulnerability’s severity. Although the system might classify $\mathit{NAT}$ with a low quantitative value, ML probabilities may necessitate assigning higher priority to severe vulnerabilities. 
In Table~\ref{tab:priority-criteria}, the first five columns calculate the priority level of the attack, while the last two columns classify the priority.
For example, if a payload is classified as an SQLi with the state $\mathit{ATT}$, and the attack’s severity is high, the system triggers a high-priority alarm that is sent to the SOC. This severity is determined based on the impact and likelihood of the attack, as classified by the CWE system. 

The Alarm Generator also gathers detailed information about the web page under attack, associating the preprocessed instance with its original URL (see Section~\ref{sec:data-prepro-detail}). This information is included in the alarm sent to the SOC team.

\begin{table}[b]\scriptsize
    \centering
    \caption{\small Priority Criteria \& Classification}
    \vspace{-2mm}
    \centering
    \resizebox{\columnwidth}{!}{%
    \begin{tabular}{@{}l|cccc||l|c@{}}
        \toprule
        \diagbox{\textbf{Severity}}{\textbf{State}} & $\mathit{NAT}$ & $\mathit{AON}$ & $\mathit{UNC}$ & $\mathit{ATT}$ & \textbf{Classification} & \textbf{Levels} \\\hline
        \textbf{Low} & 0 & 1 & 2 & 3 & \textbf{Low} & 0-2 \\
        \textbf{Medium} & 1 & 2 & 3 & 4 & \textbf{Medium} & 3 \\
        \textbf{High} & 2 & 3 & 4 & 5 & \textbf{High} & 4-5 \\
        \bottomrule
    \end{tabular}%
    }
    \label{tab:priority-criteria}
\end{table}

\subsection{Operation Modes}
The proposed Multimodal AI-based system can operate in two modes, depending on what the SOC wants to prioritise (reducing false positives or increasing the system's efficiency). Therefore, the system can alternate between the following modes:

\noindent\textbf{Preemptive}. A mode that tries to reduce the number of false positives by making all instances be evaluated by the \textit{Similarity-Based Validator} module.

\noindent\textbf{Permissive}. A mode that does not try to reduce false positives but instead focuses on trying to classify the instances considered dubious, i.e., instances classified as $\mathit{UNC}$ and $\mathit{AON}$. Instances identified as $\mathit{NAT}$ are ignored, and the ones flagged as $\mathit{ATT}$ go directly to the Alarm Generator module.


\section{Implementation \& Instantiation}
\label{sec:impl_inst}

\subsection{Implementation}
\label{sec:implementation}

We implemented our system in Python 3.11. The system consists of six modules, as illustrated in Figure~\ref{fig:approach}, and was configured to detect eight classes of attacks, the ones presented in Section~\ref{sec:wia}. Some of these modules include external tools that are interconnected using Python scripts designed to orchestrate all processes autonomously. All the modules and the heuristic framework were developed from scratch, except for the NetFlow Detector and Log Extractor modules, in which we resorted to Durão’s implementation~\citep{durao2022discovery} since our focus falls more on web injection attack classification rather than data capture. Next, we provide an overview of how the system's modules were developed.

\vspace{2mm}
\noindent
\textbf{NetFlow Detector and Log Extractor.}
The NetFlow Detector captures network traffic using Wireshark~\citep{wireshark} in PCAP format via \textit{tcpdump}~\citep{tcpdump}. Traffic is processed using the \textit{SiLK}\citep{silk} framework, using the \textit{rwp2yaf2silk} and \textit{rwcut} tools that generate CSV files with their data fields customised for ML applications.
The Log Extractor operates under the logs of the Apache web server, enhanced with the \textit{mod\_dumpio} add-on to record POST requests in addition to GET requests and customised with \textit{trace7} level for its error logs.

\vspace{2mm}
\noindent
\textbf{Data Preprocessor.}
Uses a set of Python scripts we created for filtering the logs, resorting to \texttt{re} library~\citep{re} to build the appropriate patterns, as for example to extract the user inputs and discover in them dummy values. Also, to convert metacharacters into their decimal representations, and standardise the resulting strings (user inputs) following a dictionary containing SQL commands, XSS events, tags, and substrings from various attacks, along with their corresponding tokens (see Section~\ref{sec:data-prepro-detail} for examples). 
The next modules then use the result of this standardisation.

\vspace{2mm}
\noindent
\textbf{Web Attack Classifier.}
The system uses the Doc2Vec NLP algorithm from the \emph{gensim} library~\citep{gensim} to convert the standardised input data into numerical vector form in order for ML algorithms to operate.
The system employs two supervised ML algorithms: Support Vector Machine (SVM) for binary classification and Logistic Regression for multi-classification, both from the scikit-learn library~\citep{sklearn}. These models were chosen for their superior performance over alternatives such as K-Nearest Neighbor (KNN) and Random Forest (RF), as detailed in Section~\ref{sec:ml-eva}. 
The system contains two models of Logistic Regression, each one for a group of four web attacks, namely Group 1: (CRLFI, OSCI, TI, XSS)  and Group 2: (DT, SQLi, CI, XXE).

The heuristic processes probability vectors from the ML models to determine the final state of each instance, as defined in Section~\ref{subsec:web-att-det}. Algorithms~\ref{alg:multi-class-vector} and~\ref{alg:getting-state} concretize Equations~\ref{eq:ts} and~\ref{eq:states}, respectively, for getting the Temporary State of a Probability Vector and obtaining the Consensus between Temporary States. 
Once both consensus (for Group 1 and Group 2) are obtained, i.e., their states and classes, they are compared using the rule in Equation~\ref{eq:rule}. The resulting final state and class are then passed to the next module. 

\begin{algorithm}[ht!]
\footnotesize
    \caption{Temporary State of a Probability Vector.}
    \label{alg:multi-class-vector}
    \begin{algorithmic}[1]
    \State \textbf{Input:} $vector, threshold, group$
    \State \textbf{Output:} $temp\_state$
    \State $temp\_state \gets ()$
    
    \If{$group$ \textbf{is} $1$}
         \State $crlfi, osci, ti, xss \gets$ extract\_values($vector$)
         \State $probs\_and\_classes \gets$ init\_classes($crlfi, osci, ti, xss$)
    \Else
        \State $dt, sql, ci, xxe \gets$ extract\_values($vector$)
        \State $probs\_and\_classes \gets$ init\_classes($dt, sql, ci, xxe$)
    \EndIf
    
    \State $max\_val, max\_cl \gets$ get\_max\_val\_cl($probs\_and\_classes$)
    \If{$max\_val < threshold$}
        \State $temp\_state \gets (UNDEFINED, max\_cl, max\_val)$ 
    \Else
        \State $temp\_state \gets (DEFINED, max\_cl, max\_val)$ 
    \EndIf
    \end{algorithmic}
\end{algorithm}

\begin{algorithm}[ht!]
\footnotesize
    \caption{Consensus between Temporary States}
    \label{alg:getting-state}
    \begin{algorithmic}[1]
    \State \textbf{Input:} $multi\_vec, binary\_vec, t_1, t_2, group$
    \State \textbf{Output:} $state\_and\_class$
    \State $mul\_st, mul\_cl, mul\_val \gets$ classify\_vec($multi\_vec, t_1, group$)
    \State $bin\_res, bin\_cl, bin\_val \gets$ classify\_vec($binary\_vec, t_2, group$)
    \State $state\_and\_class \gets ()$
    
    \If{is\_non\_attack($mul\_st, bin\_res$)}
        \State $best\_class \gets$ normaliz($mul\_cl, mul\_val, bin\_cl, bin\_val$)
        \State $state\_and\_class \gets (NON\_ATTACK, best\_class)$ 
    \EndIf

    \If{is\_attack\_or\_not($mul\_st, bin\_res$)} 
        \State $best\_class \gets$ get\_def\_class($mul\_st, mul\_cl, bin\_st, bin\_cl$)
        \State $state\_and\_class \gets (ATTACK\_OR\_NOT, best\_class)$ 
    \EndIf

    \If{is\_uncertain\_class($mul\_st, bin\_res, mul\_cl, bin\_cl$)}
        \State $best\_class \gets$ normaliz($mul\_cl, mul\_val, bin\_cl, bin\_val$)
        \State $state\_and\_class \gets (UNCERTAIN\_CLASS, best\_class)$ 
    \EndIf

    \If{is\_attack($mul\_st, bin\_res, mul\_cl, bin\_cl$)}
        \State $state\_and\_class \gets (ATTACK, mul\_cl)$ 
    \EndIf
    \end{algorithmic}
\end{algorithm}

\vspace{2mm}
\noindent
\textbf{Similarity-Based Validator.}
The module leverages an LSH implementation from the \emph{datasketch} library~\citep{datasketch}, with the Jaccard distance as the metric, and uses eight LSH databases, each created specifically for each attack class, parametrised for shingles of 3 tokens and 256 permutations with the goal of minimizing errors and ensuring the database query results are sufficiently accurate. 
Algorithm~\ref{alg:decide-lsh-class} describes the process of determining the attack class through LSH database queries. 
Once the best class is determined, the system applies the rule in Equation~\ref{eq:lsh-rule}, depending on the mode (Preemptive or Permissive), to decide the best state and class to forward to the Alarm Generator module.

\begin{algorithm}[ht!]
\footnotesize
    \caption{Determining the LSH Class}
    \label{alg:decide-lsh-class}
    \begin{algorithmic}[1]
    \State \textbf{Input:} $lsh\_list$, $instance$
    \State \textbf{Output:} $lsh\_class$
    \State $biggest\_jacc\_sim \gets 0$
    \State $lsh\_class \gets None$
    \ForAll{$(lsh\_key, lsh\_database, lsh\_tokens)$ \textbf{in} $lsh\_list$}
        \State $inst\_mh \gets $  create\_minhash($instance$, 3, 256)
        \State $key\_list \gets$ $lsh\_model$.query($inst\_mh$)
        \ForAll{$key$ \textbf{in} $key\_list$}
            \State $lsh\_token \gets lsh\_tokens[key]$
            \State $lsh\_mh\_tok \gets$ create\_minhash($lsh\_token$, 3, 256)
            \State $jacc\_sim \gets$ cal\_direct\_jaccard($inst\_mh$, $lsh\_mh\_tok$)
            \If{$biggest\_jacc\_sim < jacc\_sim$}
                \State $biggest\_jacc\_sim \gets jacc\_sim$
                \State $lsh\_class \gets lsh\_key$
            \EndIf
        \EndFor
    \EndFor
    \end{algorithmic}
\end{algorithm}

\vspace{2mm}
\noindent
\textbf{Alarm Generator.}
It is a Python script that aggregates information from the other modules and Table~\ref{tab:priority-criteria} to generate alarms and send them to SOC analysts. The module also includes the CWE information about the attack in the alarms, which is retrieved from a database built with the CWEs associated with attack classes supported by the system; for example, CWE-89 for SQL Injection.

\subsection{Instantiation}
\label{ssec:instantiation}

We present an instantiation of the system, including the environment in which it was deployed and the way datasets were constructed.

\subsubsection{Environment Scenario}
\label{sec:dev-sec-imp}

A controlled network was used to test scenarios and collect data.
This network allowed the use of vulnerable applications and web scanners in a safe environment, ensuring control over traffic passing through the web server.

The system was deployed in two Virtual Machines (VMs), under a host machine with 16 GB of RAM and an AMD Ryzen 5 4600H @ 3.00GHz processor, and ran a 64-bit Windows OS. One VM was configured with 4 GB RAM and 2 processors, and acted as a server, hosting the Apache Web Server, for web application hosting and storage, and the system itself. The other VM was configured with 8 GB RAM and 4 processors, and acted as a client. 

For testing, creating datasets, and evaluation purposes, in the server VM, we installed on the web server two PHP/MySQL web applications with known vulnerabilities: BWAPP~\citep{bWAPP} and DVWA\citep{dvwa}. In the client VM, we installed four web scanners: ZAP~v2.14.0~\citep{zap}, Burp~v2023.9.10~\citep{burp}, Wapiti~v3.1.3~\citep{wapiti}, and w3af~v1.6.49~\citep{w3af}, which played the role of attackers and generated traffic for the datasets.

\subsubsection{Datasets Creation and Preparation}
\label{sec:datasets}

Web scanners served to compose the datasets, and thereby they attacked \emph{separately} and \emph{isolatedly} both web applications for the eight types of injection attacks the system supports. 
\emph{Isolatedly}, we mean that we collected each flow containing only one type of injection attack, thus allowing us to process them isolatedly and so enabling efficient data processing through the Data Preprocessor by removing dummy values and standardising the payloads. \emph{Separately}, we mean that this process was made for each target application individually, in order to ensure data independence from any specific web application.
We confirmed this independence by calculating the Jaccard similarity between instances from BWAPP and DVWA. As the Data Preprocessor only retains user values after removing input fields, the similarity score indicated no dependency on the target web application, thus confirming that web scanners inject the same inputs regardless of the target.

Table~\ref{tab:instances_ratio}, columns 2 and 6, shows the number of collected instances of each class. To address class imbalance within Groups 1 and 2 of classes, we performed random under-sampling on the OSCI, SQLi, CI, TI, and XSS classes, avoiding augmentation to prevent bias from generating synthetic vectors with the SMOTE tool~\citep{smote}. 
Columns 3 and 7 show the final number of instances of each class after applying this process, and columns 4 and 8 show the ratio applied to achieve balanced datasets within each group. 

\begin{table}[t]\scriptsize
	\centering
	\caption{\small Collected and final number of instances per class and the ratio used in each one.}
    \vspace{-3mm}
	\resizebox{\columnwidth}{!}{%
		\begin{tabular}{cccc||cccc}
			\toprule
			\multirow{2}{*}{\textbf{Class}} & \textbf{Collected} & \textbf{Final} & \multirow{2}{*}{\textbf{Ratio}} & \multirow{2}{*}{\textbf{Class}} & \textbf{Collected} & \textbf{Final} & \multirow{2}{*}{\textbf{Ratio}} \\
            & \textbf{Instances} & \textbf{Instances} & & & \textbf{Instances} & \textbf{Instances} & \\\hline
            \textbf{CRLFI} & 418 & 418 & 1 & \textbf{DT} & 568 & 568 & 1 \\
            \textbf{OSCI} & 2611 & 548 & 0.21 & \textbf{SQLi} & 2766 & 553 & 0.2 \\
            \textbf{TI} & 3971 & 556 & 0.14 & \textbf{CI} & 1074 & 547 & 0.51 \\
            \textbf{XSS} & 1150 & 552 & 0.48 & \textbf{XXE} & 598 & 598 & 1 \\
            \bottomrule
		\end{tabular}%
	}
	\label{tab:instances_ratio}
    \vspace{-3mm}
\end{table}

We constructed, then, 10 datasets for the ML algorithms. For each multi-class model of each group of attack classes (Groups 1 and 2), a dataset was created containing instances from each Group's classes, with labels representing each class. This resulted in datasets of 2,074 instances for Group 1 and 2,226 instances for Group 2. 
For binary datasets, instances of the target class were labelled as 1, while other instances known by the corresponding multi-class model were labelled as 0 (e.g., the DT binary model includes DT instances as class 1, and SQLi, CI, and XXE as class 0). This setup avoided bias by ensuring each binary model has only the instances that its multi-class counterpart knows.
Models were then evaluated on datasets using a 70/30 train-test split and \textcolor{black}{10-fold cross-validation}.

\section{Experimental Evaluation}
\label{sec:evaluation}

The objective of this section is to evaluate the Multimodal AI-based system regarding the challenges of Section~\ref{ssec:challenges} and the components it comprises, namely the NLP and ML models in the Web Attack Classifier, the heuristic framework, and the Similarity-Based Validator module, both individually and within the integrated system.
Based on these aspects, we defined the following questions:
\textbf{Q1.} Can the NLP models accurately create distinguishing vectors using instances of web injection attacks?
\textbf{Q2.} Can the ML models correctly classify the vectors created by the NLP models by their type of attack?
\textbf{Q3.} Can the heuristic framework classify a vector with certainty based on the probability predictions provided by the ML models?
\textbf{Q4.} Can the heuristic framework that uses an LSH implementation perform multi-classification and improve our classification process?

\subsection{NLP Model Assessment}\label{sec:nlp-eva}
We aimed to determine whether the Doc2Vec NLP model could generate vectors capable of representing an attack by its class (to answer Q1). To do so, we varied Doc2Vec's hyperparameters for each of the 10 ML models, keeping the number of epochs to ensure consistency, and evaluated them using a set of custom metrics we created for this purpose.

\textcolor{black}{Our choice of Doc2Vec, rather than transformer-based models (e.g., BERT), for example, is motivated by the nature of the analysed data and system constraints. Web attack payloads are typically short and contain structured patterns with limited contextual depth. In such cases, lightweight embedding methods such as Doc2Vec are sufficient to capture discriminative features~\citep{pranjic2020evaluation}.
Additionally, Doc2Vec offers significantly lower computational overhead in terms of memory usage and inference latency, making it more suitable for near real-time SOC environments. Using transformer models would introduce higher higher and unnecessary computational costs without a clear proportional gain in classification performance for this type of data.
Therefore, the selection reflects a trade-off favouring efficiency and deployability while maintaining high classification accuracy.}

\begin{figure}[t]
	\centering
	\begin{minipage}{0.49\columnwidth}
		\centering
		\includegraphics[width=\linewidth]{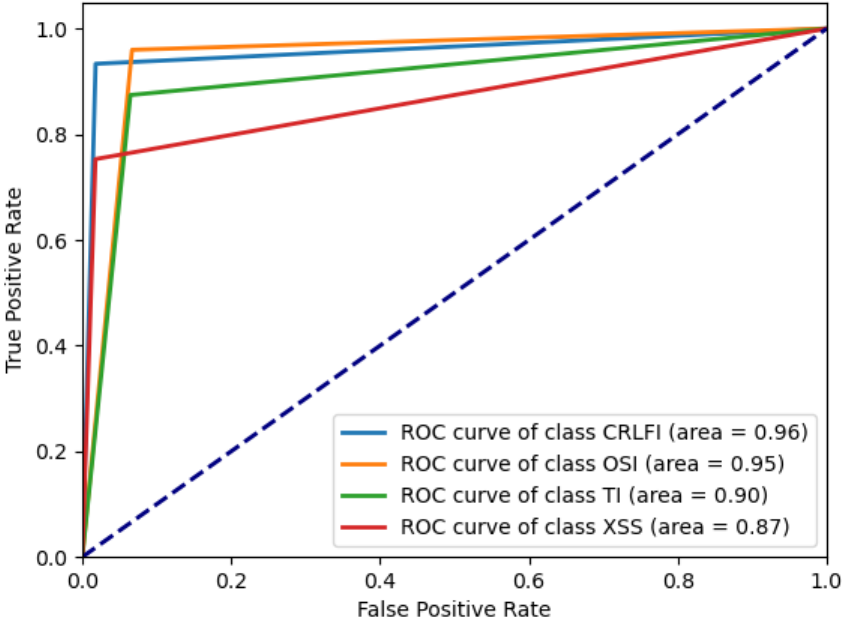}
		\subcaption{Classes of Group 1.}
		\label{fig:group1_roc}
	\end{minipage}%
	\hfill
	\begin{minipage}{0.49\columnwidth}
		\centering
		\includegraphics[width=\linewidth]{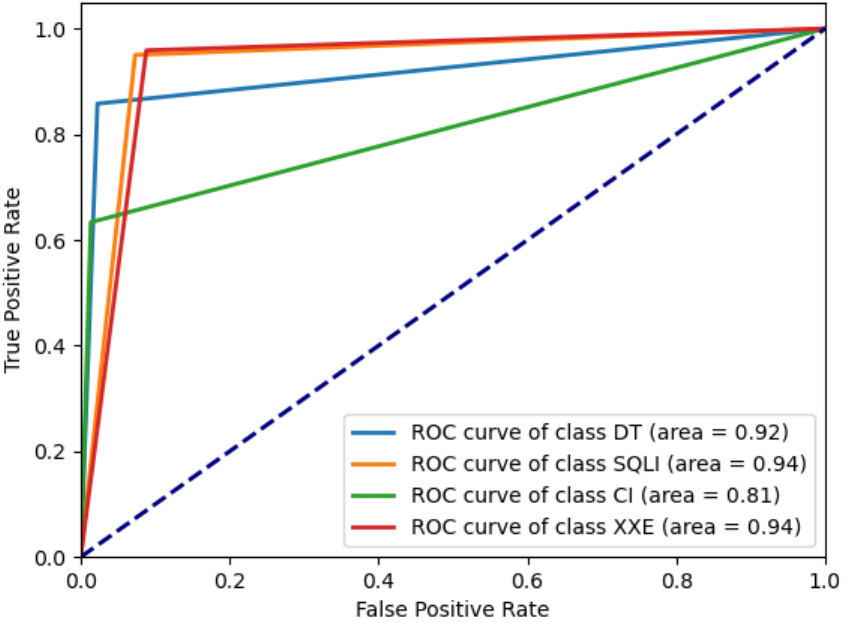}
		\subcaption{Classes of Group 2.}
		\label{fig:group2_roc}
	\end{minipage}
    \vspace{-3mm}
	\caption{ROC curves and AUC values of the classes, aggregated by the groups of multi-class models.}
    \label{fig:roc_auc}
    \vspace{-3mm}
\end{figure}

For each attack in the test set, the model generated a vector with class $c_i$ and identified the most similar vector from its training dataset. A correct classification occurred if the prediction class $c_p$ of the nearest vector matched $c_i$. The accuracy metric was calculated, in which both multi-class models achieved 0.88, and binary models ranged 0.86 to 0.98, with the CI model being the one that did not reach 0.90 accuracy.  
However, while accuracy measures the frequency of correct predictions, it does not capture false positive and negative rates, which are critical for model evaluation. We analysed the ROC (Receiver Operating Characteristic) curve and AUC (Area Under the Curve), which provide insight into false rate performance~\citep{HAIXIANG2017220}.
Figure~\ref{fig:roc_auc} displays the ROC curves and AUC for Group 1 and Group 2 of the classes. Both show ROC curves near the top-left corner, indicating high effectiveness in detecting attacks. Most AUC values for the binary models range between 0.9 and 1, confirming excellent model quality. Exceptions include the AUC for the XSS class in the Group 1 model and the CI class in the Group 2 model, which had, respectively, 0.87 and 0.81. Although they are lower than the other models, they still indicate being models with quality results, and are so reliable.

Based on these results, we conclude that for the NLP model, command and metacharacter order is generally less significant than the content of the attack. The exception is the binary model for the TI class, where the order of HTML tags and JavaScript functions may impact identification due to the TI attack’s nature. Thus, we can positively answer Q1.

\subsection{ML Model Assessment}
\label{sec:ml-eva}

This assessment evaluates the 10 ML models and addresses Q2. We tested four ML algorithms, tuning models for each class: SVM, Logistic Regression (LR), KNN Classifier (KNN), and Random Forest (RF). 
To optimise the models' hyperparameters, we used a grid search with 10-fold cross-validation on the training dataset. Each algorithm-dataset combination was evaluated, yielding a total of 40 models, each optimised with the best hyperparameters.

To prevent overfitting or underfitting, we considered each model's best $f_1$ score and used Stratified 10-Fold Cross Validation~\citep{Fon22}. Stratified K-Fold Cross Validation, unlike traditional K-Fold, maintains the class distribution in each fold, which is particularly useful for imbalanced datasets. Although our datasets were balanced, this method provides an additional safeguard against underfitting or overfitting by calculating mean Cross-Validation Accuracy $m$, Standard Deviation $std_m$, and Training Accuracy $p$.
According to Equation~\ref{eq:model-fit}, if the absolute difference between $p$ and $m$ exceeds $std_m$, the model is considered underfitted, as the training accuracy falls outside the acceptable range despite a high mean cross-validation accuracy. Otherwise, it is considered good-fitted. Conversely, we deem a standard deviation above 2\% as high, indicating overfitting if exceeded, as even small accuracy fluctuations (e.g., between 88\% and 92\%) are significant when the mean is 90\%~\citep{bates2024cross}.

\begin{equation}\label{eq:model-fit}
\small
    \begin{dcases}
		Underfit, & p \notin [m-std_m, m+std_m] \\
        Goodfit, & p \in [m-std_m, m+std_m] \\
        Overfit, & std_m >= 0.02
	\end{dcases}
\end{equation}

After this analysis and models evaluation, we identified the best model for each dataset\textcolor{black}{, i.e., the model that has the $std_m$ less than 2\%}. In cases of ties, the model with the highest $f_1$ score was selected. Table~\ref{tab:ml_results} presents the results of the top models, including $accuracy$, $precision$, $recall$, and $f_1$ score values.

\begin{table}[t]\scriptsize
    \centering
    \caption{\small Results of the best good-fitted models for every dataset.}  
    \vspace{-3mm}
    \resizebox{\columnwidth}{!}{%
    \begin{tabular}{ccccccc}
        \toprule
        Class & Group & Model & $accuracy$ & $precision$ & $recall$ & $f_1$ score \\ \hline
        multi-class & 1 & LR & 0.9197 & 0.9197 & 0.9197 & 0.9197 \\
        OSCI & 1 & SVM & 0.9502 & 0.9511 & 0.9502 & 0.9505 \\
        TI & 1 & SVM & 0.9390 & 0.9398 & 0.9390 & 0.9371 \\
        XSS & 1 & SVM & 0.9326 & 0.9319 & 0.9326 & 0.9316 \\
        CRLFI & 1 & SVM & 0.9823 & 0.9822 & 0.9823 & 0.9823 \\\hline
        multi-class & 2 & SVM & 0.9046 & 0.9104 & 0.9046 & 0.9026 \\
        DT & 2 & SVM & 0.9736 & 0.9738 & 0.9736 & 0.9737\\
        SQLI & 2 & LR & 0.9692 & 0.9690 & 0.9692 & 0.9688\\
        CI & 2 & SVM & 0.9104 & 0.9121 & 0.9104 & 0.9050\\
        XXE & 2 & SVM & 0.9325 & 0.9394 & 0.9325 & 0.9341 \\
        \bottomrule
    \end{tabular}
    }
    \label{tab:ml_results}
    \vspace{-3mm}
\end{table}

A common feature across the best models is the use of either SVM or LR algorithms. Both algorithms are inherently binary classifiers, so techniques were applied to adapt them for multi-class models. Specifically, LR uses Softmax Regression (Multinomial Logistic Regression), where the softmax function replaces the sigmoid to model class probabilities, ensuring they sum to 1. SVM, on the other hand, employs a One-vs-Rest (OvR) approach, training a separate SVM classifier for each class. During prediction, each classifier generates a confidence score, and the class with the highest score is selected.
Overall, the achieved metrics for all 10 models were greater than 0.9, with an $f_1$ score ranging from 0.9 to 0.98. These results indicate that the models correctly classify the vectors created by Doc2Vec according to their type of attack, and thus respond positively to Q2.

\subsection{Heuristic Analysis for Attack classification}\label{sec:heu-eva}

To evaluate the ability and effectiveness of the Heuristic Framework (HF) to classify instances into attack classes using the predictions of the 10 models, and so to answer Q3, we performed three experiments using HF against instances of: $i)$ known classes; $ii)$ unknown classes; $iii)$ benign inputs.   

\subsubsection{Instances of Known Classes}

In this experiment, we assessed the test set (1,304 instances) using HF configured with the best NLP and ML models identified in the previous two sections, and varying the multi-classification threshold $t_1$ between 0.25 and 0.95 and the binary threshold $t_2$ between 0.5 and 0.95, both with increments of 0.05. This yielded 150 combinations of threshold pairs to analyse the state and class returned by the heuristic on the test set.

\vspace{2mm}
\noindent\textbf{Heuristic Class Analysis.}
Table~\ref{tab:heuristic-states} shows \textcolor{black}{representative results for different ($t_1$:$t_2$) pairs, including combinations with low and high threshold values and the operating pair selected for the subsequent analyses}.
Columns display thresholds $t_1$ and $t_2$, and instance counts for states $\mathit{ATT}$ (Attack), $\mathit{UNC}$ (Uncertain Class), $\mathit{AON}$ (Attack Or Not), and $\mathit{NAT}$ (Non-Attack).

The initial rows indicate that increasing $t_2$ raises $\mathit{AON}$ counts while reducing $\mathit{ATT}$ and $\mathit{UNC}$. This is because, as shown in Figure~\ref{fig:heuristic}, a temporary state $\mathit{DEF}$ in $v_m$ and an increased $t_2$ sets $v_b$ to $UND$, resulting in $\mathit{AON}$. In the last rows, where $t_1$ is maximised, $v_m$ often assumes the $\mathit{UND}$ state, limiting final states to $\mathit{AON}$ or $\mathit{NAT}$. For example, in the $(0.95:0.95)$ pair, the 643 instances classified as $\mathit{ATT}$ occur where both $max(v_m) > t_1$ and $max(v_b) > t_2$ with matching classes.
\textcolor{black}{Among the evaluated configurations, the pair $(0.35:0.50)$ classified 1,207 instances as $\mathit{ATT}$ while producing 77 $\mathit{AON}$ and 3 $\mathit{NAT}$ classifications. 
We selected this pair as the operating configuration for the subsequent analyses because the intended operational criterion was to favor maximizing definite $\mathit{ATT}$ classifications while minimizing the less actionable $\mathit{AON}$ and $\mathit{NAT}$ ones.}
With this \textcolor{black}{configuration}, the heuristic reached a recall of 0.93, considering that $\mathit{ATT}$ are True Positives (TP), and $\mathit{AON}$ and $\mathit{NAT}$ are False Negatives (FN). 
However, recall can be higher, reaching 0.94, if we also consider the $\mathit{UNC}$ as TP, since it gives certainty that the instance is an attack, but can    not precise which class it is.

\begin{table}[t]\footnotesize\scriptsize
	\centering
	\caption{\small State results of the HF with NLP and ML models.}
    \vspace{-3mm}
	\begin{tabular}{cc|cccc}
        \toprule
		\textbf{$t_1$} & $t_2$ & $\mathit{ATT}$ & $\mathit{UNC}$ & $\mathit{AON}$ & $\mathit{NAT}$ \\\hline
		0.25 & 0.50 & 1204 & 20 & 80 & 0\\
		0.25 & 0.55 & 1194 & 18 & 92 & 0\\
		0.25 & 0.60 & 1184 & 14 & 106 & 0\\
		0.30 & 0.50 & 1199 & 27 & 78 & 0\\
		0.30 & 0.55 & 1191 & 24 & 89 & 0\\
		0.30 & 0.60 & 1180 & 18 & 106 & 0\\\hline
		\textbf{0.35} & \textbf{0.50} & \textbf{1207} & \textbf{17} & \textbf{77} & \textbf{3} \\\hline
		0.90 & 0.85 & 823 & 1 & 115 & 365\\
		0.90 & 0.90 & 790 & 1 & 122 & 391\\
		0.90 & 0.95 & 692 & 1 & 195 & 415\\
		0.95 & 0.85 & 744 & 1 & 182 &377\\
		0.95 & 0.90 & 729 & 1 & 153 & 421\\
		0.95 & 0.95 & 643 & 1 & 204 & 456 \\
        \bottomrule
	\end{tabular}
	\label{tab:heuristic-states}
\end{table}

\vspace{2mm}
\noindent\textbf{Heuristic vs Real Class Analysis.}
Through HF, an instance can be classified as an attack ($\mathit{ATT}$), but this may not guarantee correct attribution to the corresponding attack class. 
In addition to evaluating the heuristic results for the 
\textcolor{black}{selected operating} pair ($0.35:0.50$), we assessed the class predictions it outputs based on the rule: for $\mathit{ATT}$, the heuristic returns class $c_p$, while for $\mathit{UNC}$, $\mathit{AON}$, and $\mathit{NAT}$, it produces class $c_t$ (see Section~\ref{subsubsec:heu-detail}), and matched it with the real classes. Following this, we constructed the confusion matrix (Table~\ref{tab:confusion_matrix_0.35:0.5}) and calculated the metrics derived from it.

\begin{table}[t]\scriptsize
	\centering
	\caption{\small Confusion Matrix of the classes predicted by HF with the threshold pair $(0.35:0.50)$ and the real class.}
    \vspace{-3mm}
	\resizebox{\columnwidth}{!}{%
		\begin{tabular}{c|cccccccc}
            \toprule
			\diagbox{\textbf{True}}{\textbf{Predicted}} & \textbf{SQLi} & \textbf{XSS} & \textbf{DT} & \textbf{OSCI} & \textbf{CI} & \textbf{TI} & \textbf{XXE} & \textbf{CRLFI} \\
			\hline
			\textbf{SQLi} & 151 & 0 & 3 & 0 & 5 & 0 & 1 & 0 \\
			\textbf{XSS} & 0 & 151 & 0 & 3 & 0 & 9 & 0 & 7 \\
			\textbf{DT} & 3 & 0 & 184 & 0 & 3 & 0 & 0 & 0 \\
			\textbf{OSCI} & 0 & 6 & 0 & 164 & 0 & 3 & 0 & 1 \\
			\textbf{CI} & 6 & 0 & 5 & 0 & 113 & 0 & 37 & 0 \\
			\textbf{TI} & 0 & 5 & 0 & 10 & 0 & 143 & 0 & 1 \\
			\textbf{XXE} & 1 & 0 & 1 & 0 & 5 & 0 & 163 & 0\\
			\textbf{CRLFI} & 0 & 1 & 0 & 3 & 0 & 1 & 0 & 115\\
            \bottomrule
		\end{tabular}
	}
	\label{tab:confusion_matrix_0.35:0.5}
    \vspace{-3mm}
\end{table}

\begin{figure}[t]
	\centering
	\includegraphics[width=1\linewidth]{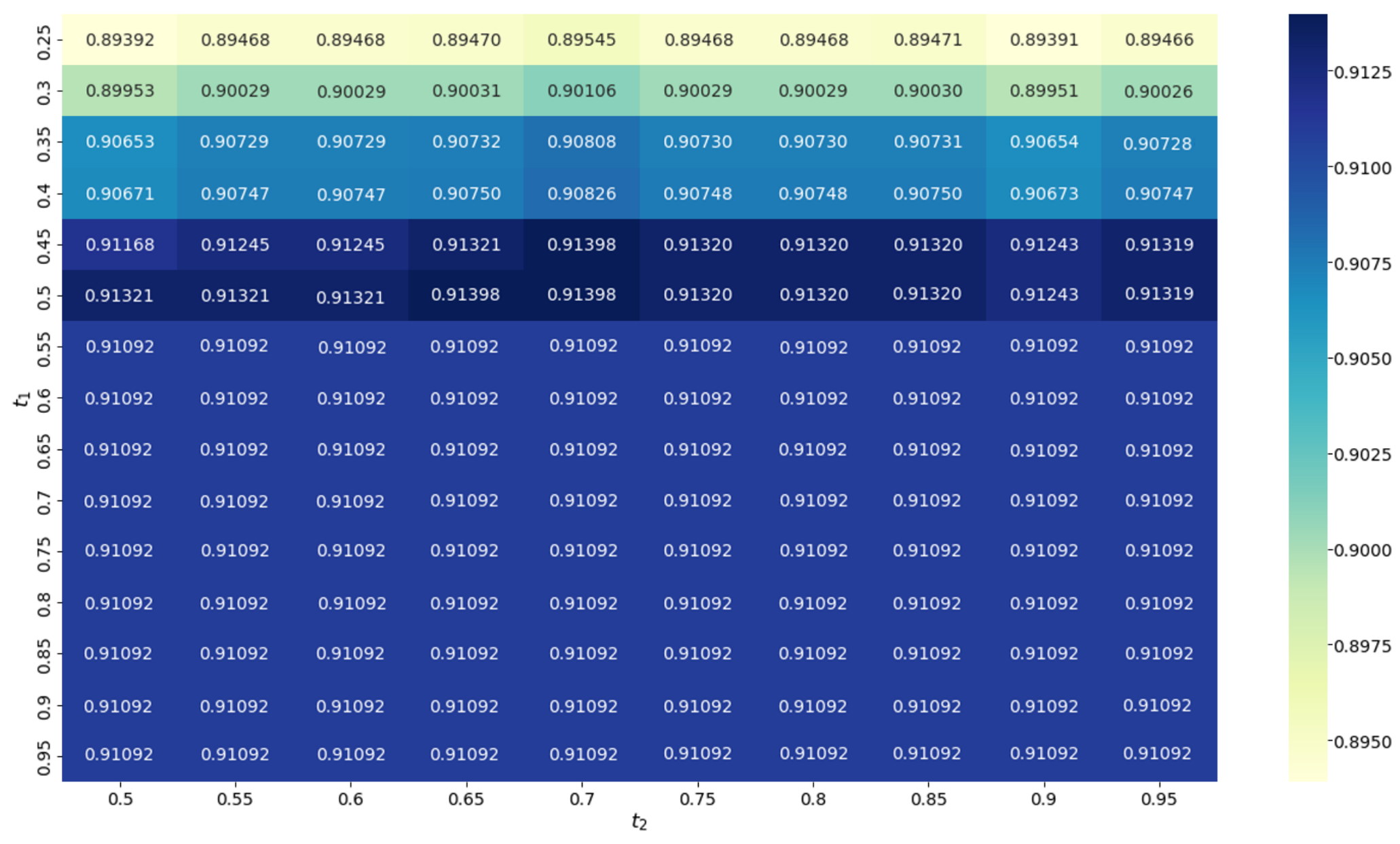}
    \vspace{-8mm}
	\caption{$weighted$ $f_1$ score Heatmap along the thresholds pairs $t_1:t_2$}
	\label{fig:heatmap-f1score}
    \vspace{-3mm}
\end{figure}

Figure~\ref{fig:heatmap-f1score} illustrates the  $weighted$ $f_1$ scores for each threshold pair.
\textcolor{black}{The selected pair ($0.35:0.50$) achieved a $weighted$ $f_1$ score of $0.90653$, whereas the highest value, $0.91398$, was obtained with ($0.45:0.70$), ($0.50:0.65$), and ($0.50:0.70$). 
Therefore, ($0.35:0.50$) was selected according to the operational objective of favoring definite $\mathit{ATT}$ classifications while limiting $\mathit{AON}$ and $\mathit{NAT}$ outcomes, not to maximize the class-level metric. 
Moreover, all evaluated combinations achieved $weighted$ $f_1$ scores above $0.89$, indicating stable classification performance across a broad range of operating thresholds. 
When applying the proposed method to other datasets or deployment environments, these thresholds should be recalibrate according to the classifier outputs, data distribution, and operational requirements, preferably using a held-out validation split or nested cross-validation.}

%

To better understand the differences between the heuristic and real classes, we calculated the Misclassification Frequency $\mathit{MF}_c$ for each class by measuring the number of false positives (FP) of each class over all FPs. 
The resulting values are presented in Table~\ref{tab:MF_c}, in which we notice that high $\mathit{MF}$ values indicate classes that are frequently misclassified, helping identify difficult-to-classify cases. 
CI is the most frequently misclassified class, often confused with XXE, possibly due to class similarity (Section~\ref{subsec:lsh-eva-ind} gives further insights about this). 
This suggests the CI class data may need more specific processing to reduce misclassification.

\begin{table}[b]\footnotesize\scriptsize
    \vspace{-3mm}
	\centering
	\caption{\small Misclassification Frequency of each class.}
    \vspace{-3mm}
	\begin{tabular}{cc||cc||cc||cc}
        \toprule
		$c$ & $\mathit{MF}_c$ & $c$ & $\mathit{MF}_c$ & $c$ & $\mathit{MF}_c$ & $c$ & $\mathit{MF}_c$ \\\hline
		DT & 0.26 & CRLFI & 0.58 & CI & 2.73 & TI & 0.76 \\
		SQL & 0.68 & OSCI & 0.63 & XXE & 0.35 & XXS & 0.88\\
        \bottomrule
	\end{tabular}
	\label{tab:MF_c}
\end{table}

In conclusion, the heuristic’s effectiveness closely matches the ML models, sometimes with deviations in attack class identification, but even so, it is capable of providing valuable insights about attacks for SOC analysis.

\subsubsection{Instances of Unknown Classes}
In this second experiment, we aim to investigate how the heuristic framework performs with instances from attack classes it does not support, i.e., its ML models and their corresponding datasets are not trained on.    
In machine learning, testing a supervised model on unfamiliar categories can introduce bias, potentially misrepresenting the system’s capabilities~\citep{lu2019}. However, since our system is intended for NIDS environments, where new kinds of injection attacks are emerging over time, its ML models will inevitably encounter unknown attack types. Therefore, it is important to see how the system behaves in such a scenario; more specifically, whether the heuristic can handle unknown attacks effectively. To do so, we compiled a dataset with 131 LDAP injection attacks, following the process in Section~\ref{sec:datasets}, and it went through the pipeline of the system.
LDAP injection targets vulnerabilities in LDAP servers by altering input parameters, often using LDAP metacharacters like wildcards (*), parentheses, and filter operators (\&, $|$, !). Attackers may use these to bypass authentication, access data without authorisation, or alter directory entries. Preventing LDAP injection involves input validation, parameterised queries, and limiting access rights.

\begin{table}[t]\footnotesize \scriptsize
	\centering
	\caption{\small LDAP Injection and Benign State results.}
    \vspace{-3mm}
    \resizebox{\columnwidth}{!}{%
	\begin{tabular}{cc|cccc|cccc}
        \cmidrule{3-10}
        \multicolumn{2}{c}{} & \multicolumn{4}{c|}{\textbf{LDAP Injection}} & \multicolumn{4}{c}{\textbf{Benign}} \\ 
        \midrule
		$t_1$ & $t_2$ & $\mathit{ATT}$ & $\mathit{UNC}$ & $\mathit{AON}$ & $\mathit{NAT}$ & $\mathit{ATT}$ & $\mathit{UNC}$ & $\mathit{AON}$ & $\mathit{NAT}$ \\
        \midrule
		0.25 & 0.50 & 45 & 74 & 1 & 11 & 0 & 12 & 86 & 108 \\
		0.25 & 0.55 & 48 & 68 & 1 & 14 & 0 & 12 & 79 & 115 \\
		0.25 & 0.60 & 52 & 62 & 1 & 16 & 0 & 12 & 75 & 119 \\\hline
		\textbf{0.35} & \textbf{0.50} & \textbf{43} & \textbf{76} & \textbf{1} & \textbf{11}  & \textbf{0} & \textbf{0} & \textbf{85} & \textbf{121} \\\hline 
		0.95 & 0.85 & 27 & 18 & 20 & 66 & 0 & 0 & 80 & 126 \\
		0.95 & 0.90 & 31 & 5 & 20 & 76 & 0 & 0 & 73 & 133 \\
		0.95 & 0.95 & 14 & 3 & 20 & 94  & 0 & 0 & 69 & 137 \\
        \bottomrule
	\end{tabular}%
    }
	\label{tab:ldap-benign-states}
    \vspace{-3mm}
\end{table}

The results of the heuristic classification of the dataset based on the ML models are shown in Table~\ref{tab:ldap-benign-states}, columns 3 to 6, with the first and last three threshold pairs with $t_1$ equal to 0.25 and 0.95, and the \textcolor{black}{selected operating} threshold ($0.35:0.50$).
%
Our primary expectation was that most instances would classify as $\mathit{NAT}$, reflecting that they were unknown attacks. However, the results provided further insight into how the system interacts with unfamiliar instances. 
Increasing $t_1$ led to fewer $\mathit{ATT}$ instances and more $\mathit{NAT}$ classifications. Similarly, raising $t_2$ reduced $\mathit{ATT}$ and $\mathit{UNC}$ counts while increasing $\mathit{AON}$ or $\mathit{NAT}$ classifications. Unexpectedly, many instances under the \textcolor{black}{selected} pair $(0.35:0.50)$ fell into $\mathit{UNC}$ (77 out of 131), which we initially assumed indicated attacks of known classes, and many other instances were classified as $\mathit{ATT}$ (46 out of 131). 
However, upon analysis, it became clear that these states, in particular $\mathit{UNC}$, can also include unknown classes. In the specific case of LDAP injection attacks, they contain metacharacters shared with other injection types.

This finding suggests that the system can detect unknown web injection attacks, as they contain metacharacters that it knows, filtering them into the $\mathit{ATT}$ or $\mathit{UNC}$ states. 
Considering these states as TP and other states as FN, recall is $0.91$, a respectable value, given the model was untrained on LDAP injections. 
While this evaluation may not fully reflect the model’s capabilities in correctly identifying the attack class, the results in such scenarios are valuable for SOC teams since they indicate the existence of attacks.

\subsubsection{Instances of Benign Inputs}
This last experiment aims to find out how the heuristic performs when managing benign inputs, and thereby checks its ability to distinguish attacks from non-attacks. 

As the system to collect inputs, they would only reach the Log Extractor module if embedded in anomalous traffic. This means that benign inputs would appear suspicious, containing malicious metacharacters, which is unlikely, as inputs with such characters are often treated as potential vulnerabilities~\citep{shariar2012}.
Nonetheless, we simulated a scenario where benign inputs reach the Web Attack Classifier module by deactivating the Data Preprocessor’s benign metacharacter filter, allowing benign characters like \texttt{\%40} (\texttt{@}) to pass through. We created 206 benign inputs, including some extreme cases with SQL commands and JavaScript functions.

Table~\ref{tab:ldap-benign-states}, columns 7 to 10, display heuristic results for previously used threshold pairs. In threshold pairs above the \textcolor{black}{highlighted} pair, most instances were classified as $\mathit{NAT}$ or $\mathit{AON}$, with only 12 out of 206 classified as $\mathit{UNC}$, those that contained SQL commands or JavaScript. Notably, the \textcolor{black}{selected operational} threshold pair ($0.35:0.50$) did not classify any benign instance as $\mathit{ATT}$ or $\mathit{UNC}$, indicating that the system can avoid misclassifying benign inputs even in extreme cases.

In conclusion, the system demonstrates strong effectiveness with benign inputs, predominantly classifying them as $\mathit{AON}$ or $\mathit{NAT}$. This reduces their priority for the SOC team and minimizes false positives.
Furthermore, we can conclude that the HF using the results of the ML models classifies attacks correctly with a high recall and accuracy, which, therefore, leads us to answer Q3 positively.

\subsection{Heuristic Analysis for Attack Validation \& Misclassification}
\label{sec:lsh-eva}

In this section, we aim to answer Q4 by evaluating the effectiveness of the Heuristic framework in classifying attacks based on the LSH model it comprises. More specifically, we aim to evaluate the Similarity-Based Validator module in validating classifications made by the Web Attack Classifier and identifying misclassifications made by it, thus reducing potential bias and misclassification. 
The assessment was conducted in two phases. First, we evaluated the efficacy of the LSH databases independently by comparing the framework output class with the actual class. Second, we assessed the module as a whole to determine if it improves classification accuracy according to the rule in Equation~\ref{eq:lsh-rule}, thereby validating the classifications made by the Web Attack Classifier and identifying any deviations in these.

\subsubsection{Comparing LSH Classification with Real Classes}\label{subsec:lsh-eva-ind}

With the goal of evaluating the efficacy of the Similarity-Based Validator module when it works standalone, the 1,304 instances of the test set were input directly to the module, thus allowing it to return a class independently of the Web Attack Classifier. The LSH databases were built with threshold values ranging from 0.5 to 0.95. Each test used a uniform threshold across all databases. 

Tables~\ref{tab:confusion_matrix_lsh} and~\ref{tab:lsh_results} present the confusion matrix and the metrics, respectively, obtained from the performed tests.
The confusion matrix includes a $NONE$ class for instances with no similar candidates across LSH databases. This outcome is unique to LSH, differing from supervised ML models that always assign a label. Instances classified as $NONE$ are distinct from \textit{benign} because the LSH framework can only confirm that an instance does not match any known class. 

\begin{table}[t]\footnotesize
    \centering
    \caption{\small Confusion Matrix of the classes predicted by the LSH framework and the real class.}
    \vspace{-3mm}
    \resizebox{\columnwidth}{!}{%
    \begin{tabular}{c|ccccccccc}
        \toprule
        \diagbox{\textbf{True}}{\textbf{Predicted}} & \textbf{SQLi} & \textbf{XSS} & \textbf{DT} & \textbf{OSCI} & \textbf{CI} & \textbf{XXE} & \textbf{TI} & \textbf{CRLFI} & \textbf{NONE}\\
        \midrule
        \textbf{SQL} & 147 & 0 & 9 & 0 & 0 & 0 & 0 & 0 & 4\\
        \textbf{XSS} & 1 & 112 & 4 & 1 & 0 & 0 & 37 & 12 & 0\\
        \textbf{DT} & 0 & 0 & 188 & 0 & 0 & 0 & 0 & 0 & 2\\
        \textbf{OSCI} & 0 & 0 & 2 & 166 & 0 & 0 & 1 & 2 & 3\\
        \textbf{CI} & 0 & 0 & 10 & 0 & 148 & 0 & 0 & 0 & 3\\
        \textbf{XXE} & 0 & 1 & 0 & 0 & 134 & 30 & 0 & 0 & 3\\
        \textbf{TI} & 0 & 3 & 0 & 15 & 0 & 0 & 138 & 2 & 1\\
        \textbf{CRLFI} & 0 & 1 & 0 & 0 & 3 & 0 & 0 & 116 & 0\\
        \textbf{NONE} & 0 & 0 & 0 & 0 & 0 & 0 & 0 & 0 & 0\\
        \bottomrule
    \end{tabular}
    }
    \label{tab:confusion_matrix_lsh}
\end{table}

\begin{table}[t]\footnotesize
    \centering
    \caption{\small Metrics of LSH through the 1,304 instances.}
    \vspace{-3mm}
    \resizebox{\columnwidth}{!}{%
    \begin{tabular}{l|cccc}
        \toprule
        \textbf{LSH Framework} & $accuracy$ & $precision$ & $recall$ & $f_1$ score \\ \hline
        Standalone & 0.8014 & 0.8672 & 0.8014 & 0.7848 \\
        Standalone CI-XXE & 0.9076 & 0.9246 & 0.9076 & 0.9116 \\
        \bottomrule
    \end{tabular}%
    }
    \label{tab:lsh_results}
    \vspace{-3mm}
\end{table}

Overall, the results were promising, with the framework identifying 1,045 of the 1,304 attacks, thus achieving a precision of 0.87 (Standalone row of Table~\ref{tab:lsh_results}). 
The confusion matrix shows that the module achieved more TP in some classes (e.g., DT) and fewer in others (e.g., SQLi). In particular, 134 out of 170 XXE instances were misclassified as CI, likely due to the high Jaccard similarity between XXE and CI vectors. 
Such misclassification impacted the resulting metrics, reaching a recall of 0.80 and an $f1$ score of 0.78. 

Listings~\ref{lst:ssci-processing-lsh} and~\ref{lst:xxe-processing-lsh} show instances of CI and XXE attacks from their original format (performed by the attacker) when the data preprocessing is performed and the resulting shingles are built.
As we can observe, XXE and CI share many common metacharacters, resulting in high similarity values ($jacc = 0.87$), despite different attack types.

To address this, we created the CI-XXE dataset by merging the instances of both CI and XXE datasets, and labelled them as CI-XXE class. The dataset contains 331 instances, and we evaluated it to verify whether the number of false positives and negatives is reduced. The framework correctly classified 312 out of 331 instances in the new CI-XXE class, more 134 correct classifications than when these classes are separated.
Such improvement increased the metrics, with an enhancement in $f1$ score of 13\%, reaching 0.91 (Standalone CI-XXE row of Table~\ref{tab:lsh_results}).

\begin{lstlisting}[mathescape, breaklines, numbers=none, caption={A CI injection attack processing example, from its original user input to the result in the log to the data preprocessing and finally the resulting shingles.}, label={lst:ssci-processing-lsh}]      
1'+(function(){if(typeof fk7am === "undefined"){var a = new Date();do{var b = new Date();{while(b-a < 20000); fk7am = 1;}}())+'
                                $\downarrow$
1%27%2b%28function%28%29%7bif%28typeof%20fk7am%3d%3d%3d%22undefined%22%29%7bvar%20a%3dnew%20Date%28%29%3bdo%7bvar%20b%3dnew%20Date%28%29%3b%7dwhile%28b%2da%3c20000%29%3bfk7am%3d1%3b%7d%7d%28%29%29%2b%27
                                $\downarrow$
NUM%27%2B%28FUN%28%29%7BIFS%28%TYP%20TOK%3D%3D%3D%22TOK%22%29%7BTOK%20A%3DNEW%20DAT%28%29%3BDOC%7BVAR%20%3DNEW%20%DAT%28%29%3B%7DWHL%28B%2DTOK%3CNUM%29%3BTOK%3DNUM%3B%7D%7D%28%29%29%2b%27
                                $\downarrow$
['OK%', 'B%7', '%2B', '2%2', 'DNU', '20%', 'B%2', '8%T', 'CNU', 'DTO', 'A%3', '3DN', 'D%3', 'BIF', '2B%', '9%7', 'DNE', '%3D', '2b%', 'OC%', 'DAT', '27%', '0TO', 'BTO', 'DOC', 'K%2', '3B%', 'C%7', '7BI', 'W%2', '28F', 'FS%', 'R%2', '%20', '0%3', 'T%2', '22T', 'BVA', 'IFS', '8FU', '28%', '7BT', '29%', '%27', '20T', 'HL%', '%2D', 'D%7', 'VAR', 'TOK', '20D', '3D%', '%28', 'AR%', '0%D', '%3B', 'WHL', '%7D', 'M%3', '7D%', '20A', 'AT%', 'NUM', '2DT', '9%2', '8B%', '%3C', '0DA', 'NEW', 'UM%', '%22', '%DA', '3CN', 'YP%', '%2b', 'P%2', '9%3', '%7B', 'DWH', '28B', '3BT', 'L%2', 'UN%', 'D%2', 'N%2', '7BV', '%29', '%TY', '3BD', 'BDO', 'K%3', 'TYP', '2TO', '0A%', '7%2', 'M%2', 'EW%', '8%2', 'S%2', 'FUN', '7DW', '22%']
\end{lstlisting}

\begin{lstlisting}[mathescape, breaklines, numbers=none, caption={A XXE injection attack processing example, from its original user input to the result in the log to the data preprocessing and finally the resulting shingles.}, label={lst:xxe-processing-lsh}]      
<?xml version="1.0"?><!DOCTYPE foo SYSTEM "https://wapiti3.ovh/dtd/la4002/243/616374696f6e/javalin.dtd"><foo>&trick;</foo>
                                $\downarrow$
%3C%3Fxml%20version%3D%221.0%22%3F%3E%3C%21DOCTYPE%20foo%20SYSTEM%20%22https%3A%2F%2Fwapiti3.ovh%2Fdtd%2Fla4002%2F243%2F616374696f6e%2Fjavalin.dtd%22%3E%3Cfoo%3E%26trick%3B%3C%2Ffoo%3E
                                $\downarrow$
%3C%3FTOK%20VER%3D%22NUM%22%3F%3E%3C%21TYD%20TOK%SYS%20%22HTS%3A%2F%2FTOK%2ETOK%2FTOK%2FTOK%2FNUM%2FTOK%TOK%ETOK%22%3E%3CTOK%3E%26TOK%3B%3C%2FTOK%3E
                                $\downarrow$
['VER', 'NUM', '3CT', '3C%', '2NU', '%3D', '%20', '6TO', '3E%', 'E%3', 'UM%', 'S%3', '%3E', '%3C', 'B%3', 'C%2', 'S%2', 'TYD', '1TY', '%21', 'TS%', '3FT', '0TO', '%2F', 'TOK', 'F%3', '0VE', '%26', '0%2', 'E%2', '2FN', 'K%T', '3F%', '2F%', 'K%S', 'FTO', 'OK%', 'K%E', '3A%', '2ET', '26T', '2%3', 'YD%', 'CTO', '22N', 'A%2', '2FT', 'ER%', '2HT', 'R%3', '%3F', '3D%', 'M%2', 'HTS', 'F%2', 'YS%', '%SY', 'K%3', '%2E', '21T', '20T', '22%', '%TO', '20V', 'D%2', '%3A', '3B%', 'SYS', '22H', 'C%3', '20%', '%ET', '%22', 'ETO', '%3B', 'FNU', 'K%2']
\end{lstlisting}

\subsubsection{Validating Classifications}
\label{sec:eval-preemptivevspermissive}

We aim to evaluate the efficacy of the Similarity-Based Validator module when included in the system, and its ability to validate attacks discovered by the Web classifier module and flag misclassifications made by this module. Specifically, the Similarity-Based Validator needs to meet two primary objectives: i) Detect misclassifications made by the Web Attack Classifier when an instance is in the $\mathit{ATT}$ state; ii) Aid the system in determining if instances classified as $\mathit{UNC}$ and $\mathit{NAT}$ could be attacks. 
Objective i) means that the module operates in Preemptive Mode, while ii) it acts in Permissive Mode.
This evaluation was conducted both with and without the CI-XXE class, balancing information retention with improved classification efficacy.

\vspace{2mm}
\noindent\textbf{Preemptive Mode.}
In pursuit of objective i), we used the Web Attack Classifier’s classification results with the heuristic and threshold pair $(0.35:0.50)$, resulting in 74 misclassifications in the correct class out of 1,207 $\mathit{ATT}$ instances. The first two rows of Table~\ref{tab:lsh_heuristic_attack} show the classifications made by the Similarity-Based Validator module, when CI and XXE are treated as separate classes.

\begin{table}[b] \footnotesize\scriptsize
    \centering
    \caption{\small Results of the Similarity-Based Validator when operating in Preemptive and Permissive modes, and analysing the instances classified by the Web Attack Classifier.}
    \vspace{-3mm}
    \resizebox{\columnwidth}{!}{%
    \begin{tabular}{cc|cc|c}
        \cmidrule{2-5}
        & \textbf{Web Att} & \multicolumn{3}{c}{\textbf{Similarity-Based Validator}} \\ \cline{3-5}
        & \textbf{Classifier} & correctly classified & misclassified & $accuracy$\\ 
        \midrule
        \multirow{4}{*}{\rotatebox[origin=c]{90}{%
  {\centering \textbf{\footnotesize Preempt.}}}}
        & 1,133 & 1,093 & 40 & \multirow{2}{*}{0.912}\\
        & 74 & 8 & 66 & \\ \cline{2-5}
        & 1,133 & 1,109 & 24 & \multirow{2}{*}{0.951}\\
        & 74 & 40 & 34 & \\\hline
        \multirow{4}{*}{\rotatebox[origin=c]{90}{%
  {\centering \textbf{\footnotesize Permiss.}}}}
        & 50 & 9 & 41 & \multirow{2}{*}{0.171}\\
        & 44 & 8 & 36 & \\ \cline{2-5}
        & 50 & 15 & 35 & \multirow{2}{*}{0.242}\\
        & 44 & 9 & 35 & \\
        \bottomrule
    \end{tabular}%
    }
    \label{tab:lsh_heuristic_attack}
    \vspace{-3mm}
\end{table}

While the module was expected to confirm all 1,133 instances correctly classified by the Web Attack Classifier, this was not fully achieved. Out of these, 1,093 were correctly classified by the module, while 40 were misclassified, resulting in an accuracy of 0.965. For the 74 instances classified as $\mathit{ATT}$ by the Web Attack Classifier but with incorrect classes, only 8 were correctly classified by the Validator, thus identifying these as misclassifications from the first module. The Validator also misclassified the remaining 66 instances (out of 74). Overall, the achieved accuracy for the 1,101 correct classifications was 0.91. 
To understand these misclassifications from both modules, we evaluated the scenario with CI and XXE merged as a single class, referred to as CI-XXE. In cases where the actual class is XXE or CI, and the module classifies it as CI-XXE, it is considered correct. 
Table~\ref{tab:lsh_heuristic_attack}, rows 3 and 4, present the results.
It is visible that CI and XXE classes, when merged, improve the final results. Around 50\% of the previously misclassified instances by the module had their class correctly identified, reaching 1,149 correct instances and thus surpassing the 1,133 correct classified instances of the Web Attack Classifier. 
In particular, 32 of the 74 initial misclassifications involved CI misclassified as XXE; merging these classes allowed the module to identify these cases correctly.
Overall, the module achieved an accuracy of 0.95.

\vspace{2mm}
\noindent\textbf{Permissive Mode.}
To evaluate the ii) objective, we analyzed instances classified as $\mathit{UNC}$ and $\mathit{AON}$ by the Web Attack Classifier, using the same threshold pair ($0.35:0.5$) as the prior evaluation. Of the 94 instances in these states (17 $\mathit{UNC}$, 77 $\mathit{AON}$), 44 were misclassified -- 7 from $\mathit{UNC}$ and 37 from $\mathit{AON}$ -- which is reasonable given the lower confidence of these states compared to $\mathit{ATT}$. 

Table~\ref{tab:lsh_heuristic_attack}, rows 5 and 6, detail how the Similarity-Based Validator classified these instances and the resulting improvement.
The module demonstrated difficulty in correctly classifying these instances, accurately identifying only 17 out of 94 instances, which included 8 of the 44 misclassified instances from the previous module.
This limited improvement did not sufficiently offset the lower accuracy when the Web Attack Classifier made correct classifications. When treating CI and XXE as a single class, the results improved slightly, as shown in the last two rows of the table, with accuracy increasing from 0.17 to 0.24.

In conclusion, although the Similarity-Based Validator, when operating in permissive mode, presented deficits in correctly determining attacks and their classes, it improves the classification initially made by the Web Attack Classifier, as it can specify which class it belongs to. 
In the Preemptive mode, notably, the module can validate the results of the Web Attack Classifier and identify misclassifications made by this module. Such identification is valuable because it reduces the number of false positives, thereby helping SOC teams make informed decisions. Furthermore, these results align with those evaluated against the real classes. Therefore, we can conclude that the LSH model included in the proposed system plays an important role in the detection and validation of web injection attacks.
\textcolor{black}{But at the same time, in its current implementation, it should be used in Preemptive mode. The Permissive mode should be interpreted as an exploratory configuration designed to analyse uncertain instances, and therefore, is not recommended for standalone deployment.}


\subsection{\textcolor{black}{Baseline Assessments}}

\subsubsection{ModSecurity Baseline Assessment}

\textcolor{black}{As a practical rule-based baseline, we evaluated ModSecurity~\cite{modsec}, a widely used Web Application Firewall (WAF) renowned for its ability to defend against web application attacks on HTTP traffic, acting as a DPI system with pattern-based rules. ModeSecurity was configured with the OWASP Core Rule Set at its default Paranoia Level 1 (PL1), a configuration intended to provide baseline protection while limiting false positives. 
Among the 4 scanners, we opted to use Burp~\cite{burp} since it is the most powerful of them. 
We run Burp for SQLi attacks against BWAPP and with ModSecurity enabled for attack detection. 
Following the attack, we collected and analysed the logs, explicitly identifying the attacks detected by ModSecurity and those it failed to recognise.
Burp generated 901 malicious SQLi inputs against BWAPP, of which ModSecurity failed to detect 109, resulting in a recall of $87.9\%$.
This result provides a practical reference for the performance of a widely deployed rule-based WAF, which is lower than the recall our solution achieved for SQLi -- $96.9\%$ (Table~\ref{tab:ml_results}). 
}

\subsubsection{Generative AI-Based Baseline Assessment}
\textcolor{black}{To investigate whether recent generative AI models could be used as attack classifiers, we evaluated ChatGPT-4o, ChatGPT-3.5, and Gemini-1.5 using a methodology inspired by Bae \textit{et. al} \cite{electronics13132657}. A dataset containing 100 randomly selected instances of each attack class was submitted to the models using different prompting strategies, including concise, incentive-based, and step-by-step prompts. The predictions were then analysed using confusion matrices and weighted F1-scores.
The results in Table~\ref{tab:llms-results} showed limited effectiveness in all models evaluated. ChatGPT-4o achieved the best overall performance, with an average $weighted \text{ }  f_1$ of approximately 0.23, while Gemini-1.5 and ChatGPT-3.5 performed worse. A detailed analysis revealed that the models frequently produced incorrect classifications, generated unsupported attack categories, or hallucinated information not present in the original payloads.}

\begin{table}[htb]\scriptsize\footnotesize
    \centering
        \caption{Models F1 scores for each prompt type.}
        \vspace{-3mm}
        \begin{tabular}{c|c|c|c}
        \hline
        & \textbf{Concise} & \textbf{Incentive} & \textbf{Step-by-Step} \\
            \textbf{Model} & $weighted \text{ } f_1$ & $weighted \text{ } f_1$ & $weighted \text{ } f_1$\\\hline
            ChatGPT-4o & 0.210 & 0.210 & 0.282 \\
            Gemini-1.5 & 0.09 & 0.121 & 0.131\\
            ChatGPT-3.5 & 0.03 & 0.03 & 0.09 \\\hline
        \end{tabular}
    \label{tab:llms-results}
\end{table}

\textcolor{black}{These findings suggest that, in their current form, general-purpose generative AI models are not sufficiently reliable for multiclass web injection attack classification. In contrast, the proposed Doc2Vec and ML-based approach achieved substantially higher classification performance ($weighted \text{ }  f_1$ of $0.91$), supporting the use of specialised models tailored to the characteristics of web attack payloads.}

\section{\textcolor{black}{Final Remarks and Real-World Applicability}}
\label{sec:real-world}

\textcolor{black}{This work focuses on evaluating the effectiveness of detection and classification, rather than characteristics of system performance such as latency, throughput, or memory usage. Although these aspects are critical for production deployment, we prioritised the solution's effectiveness rather than its efficiency. 
The experimental setup was conducted in a controlled environment, with the primary goal of validating the feasibility and effectiveness of the proposed multimodal architecture rather than characterising production-level traffic distributions. As such, we left the efficiency evaluation for future work, which should include performance benchmarking under realistic traffic loads.
%
%
}

\textcolor{black}{In realistic deployments, benign traffic may exhibit heterogeneous patterns due to diverse application behaviours (e.g., APIs, automated scripts, or monitoring tools), potentially leading to multiple benign smaller clusters. This could increase the number of candidate NetFlows flagged for further inspection.
However, the system is designed to handle such scenarios gracefully. Since clustering acts only as a pre-filtering mechanism, an increase in candidate NetFlows results primarily in a higher processing load rather than incorrect attack classification. Downstream modules provide progressively stricter validation, mitigating false positives introduced at earlier stages.
Therefore, the system degrades in a controlled manner, trading computational efficiency for broader inspection coverage, while preserving classification accuracy.}

\section{Related Work}
\label{sec-background-related}

This section provides 
the main related work in the detection of network attacks, using NetFlows and ML.

\vspace{2mm}
\noindent\textbf{Intrusion Detection and Network Flow.}
\label{sec-netflows-detection}
Anomaly detection is typically approached through \textit{Deep Packet Inspection} (DPI) or \textit{NetFlow-based methods}.  DPI inspects packet content for malicious signatures, identifying specific threats and applications responsible for anomalies~\citep{maghraby2017}. Web application firewalls (WAF) can be seen as a particular DPI for web attacks, as they also inspect payloads~\citep{Gallego:25,Shahid:22}.
Because of this feature, DPI creates a bottleneck on networks with a high volume of traffic and can lead to a denial-of-service~\citep{cisco_vuln}. Moreover, its effectiveness is limited for encrypted traffic, as DPI depends on proxies to decrypt data, which introduces legal and security risks like data exposure~\citep{jarmoc2012ssl}.

On the other hand, NetFlow technology aggregates packets with common properties (e.g., IPs, port numbers) to provide efficient, passive monitoring capabilities~\citep{cisco-netflows,Estan2004,7509765}.
Such aggregation enables NetFlows analysis to surpass the proxy requirement by inspecting only the lower TPC/IP network layers, ensuring high-speed and large-volume networks while mitigating risks associated with DPI. It is suitable for modern encrypted network environments~\citep{Leevy2021}, offering privacy-conscious detection, including botnet command and control (C\&C)~\citep{Subramaniam2021}.
%

\vspace{2mm}
\noindent\textbf{AI for Attack Detection and Classification.}
\label{sec-ml}
Artificial Intelligence (AI) is a broad area that encompasses ML and NLP approaches to achieve its intended goals~\citep{ElNaqa2015}. 
Unsupervised and supervised ML algorithms are the most applied, and we focus on them in this paper. 
Unsupervised ML is used on a large unlabeled dataset to find a structure in unlabelled data and classify it without external validation. This technique is very effective in discovering unseen structures in the data by aggregating similar data into the same group -- cluster --, leaving the unseen data isolated in a cluster. 
LSH is considered a specific method of clustering used for big data tasks. 
Its primary goal is to handle the computational challenges associated with finding the closest data points to a given query point, a task that becomes increasingly difficult as the number of dimensions grows.
LSH enables efficient similarity checks by grouping similar data points, with distance metrics enhancing anomaly detection through clustering~\citep{le2017sonar, Qi2021lsh, bolboac2020}.
On the other hand, supervised ML relies on a dataset with instances labelled into a class/category, then trains a model to understand the pattern behind the labels and predict the class of an unseen sample. Some algorithms used for classification are, for example, Support Vector Machine (SVM)~\citep{svm} and Logistic Regression~\citep{EDGAR201795}. 
NLP, in turn, explores how machines comprehend written and spoken language, including critical application areas, like Text Processing, Syntactic, Semantic and Sentiment Analysis~\citep{Min2023}. NLP is commonly used for text classification, which involves processing the text and capturing its semantics and context to generate a unique vector representation, such as with Doc2Vec. Supervised ML models, for example, can then use this vector for classification problems.

Various works have used AI to detect attacks in network traffic, including inspecting payloads. 
Sarjan \textit{et al.}~\citep{mohanad2021} demonstrated that supervised ML can detect intrusions in inspecting NetFlows. Despite showing promising results, it presents the risk of overfitting, as the labelled data may not adequately represent unseen malicious NetFlows.  
In contrast, unsupervised ML can detect anomalous traffic with NetFlow analysis, as shown by Durão \textit{et al.}~\citep{durao2022discovery}. Their evaluation found that the source IP address differs significantly between benign and malicious traffic.
Das \textit{et al.}~\citep{das2020} proposed NLPIDS that combines NLP with ensemble-based ML to identify anomalous HTTP requests.
LSH has also been applied in anomaly detection, specifically in Industrial 4.0 environments~\citep{Qi2021lsh, bolboac2020}. 
To detect web attacks by payload inspection, 
Seyyar \textit{et al.}~\citep{seyyar2022nlp} uses deep learning, and Kar \textit{et al.}~\citep{kar2016sqligot} applies SVM, both to detect SQLi attacks on HTTP traffic. Irungu \textit{et al.}~\citep{Irungu:23} also detect SQLi attacks on HTTP by measuring metadata of SQL queries (e.g., length) and network traffic (e.g., latency), and feed supervised ML with this data to understand which model performs better.   
Gallego \textit{et al.}~\citep{Gallego:25} also evaluated supervised models over the HTTP traffic inspected by the ModSecurity WAF.  
Vadhil \textit{et al.}~\citep{Vadhil:24} studied which supervised ML model was better suited to the classification of SQLi, XSS and brute force attacks over the CIC-IDS-2017, a well-known and old dataset. The results apparently were good, but were completely biased by the use of an unbalanced dataset containing 98.7\% of benign instances and 1.2\% of those attacks. 
Deep learning models also have been used for network attack detection purposes~\citep{Yi:23}. For example, Zhou \textit{et al.}~\citep{Zhou:25} used deep learning models at 
character-level 
to classify twelve classes of attacks, employing the CSIC 2010 dataset for binary classification and the SR-BH 2020 dataset for multi-class classification.  

\textcolor{black}{These works demonstrate the use of NLP, ML, and deep learning for binary anomaly detection or multiclass web attack classification. 
Our approach differs by combining Doc2Vec-based representations with parallel binary and multiclass classifiers, an explicit consensus heuristic that assigns (un)certainty states ($\mathit{ATT}$, $\mathit{UNC}$, $\mathit{AON}$, and $\mathit{NAT}$), and an independent LSH-based stage for validating the predicted attack class.}
Regarding LSH, Durão \textit{et al.}~\citep{durao2022discovery} showed that it can be valuable for detecting web application attacks; however, their work only utilised a single LSH database to validate the attack. 
Our goal is to take this a step further by partitioning the database into multiple entities and devising a classification system for injection attacks.

\section{Conclusion}
\label{sec:conclusion}
This paper introduced a system for enhancing web application security and SOC teams' tasks by detecting and classifying web injection attacks, providing certainty. Using unsupervised ML, the system identifies suspicious HTTP(S) NetFlows and analyzes payloads through NLP, supervised ML, and a similarity search engine to detect and categorize attacks.
The system's approach detects anomalous NetFlows, preprocesses their suspicious payloads from web server logs, converting them into numeric vectors, through NLP. Logistic Regression, SVM, and a custom heuristic classify these vectors while generating certainty states.
An LSH framework is used to validate ML results, including the removal of false positives and negatives. Lastly, for detected attacks, alarms are generated for SOC teams.
The system was implemented and configured to detect eight types of web injection attacks. It achieved an $f_1$ score of 0.91 and an accuracy of 0.95, with the heuristic enabling uncertain states for misclassified instances to help SOC teams reduce false positives. The system assigns priority to each classified attack based on severity and certainty.

\section*{Acknowledgments}
This work was partially supported by P2030 through project I2DT, ref. COMPETE2030-FEDER-00389100, an ITEA4 European project (ref. 22025), and by FCT through the LASIGE Research Unit, ref. UID/00408/2025, DOI \url{https://doi.org/10.54499/UID/00408/2025}.

\bibliographystyle{elsarticle-harv} 
\bibliography{bib}

\LONG{

} 

%
\subsection*{  } 
\setlength\intextsep{0pt} 
\begin{wrapfigure}{l}{0.13\textwidth}
    \centering
    \includegraphics[width=0.15\textwidth]{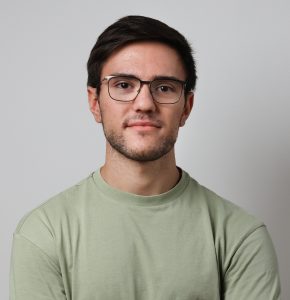}
\end{wrapfigure}
\noindent \textbf{Rodrigo Branco} is a CyberSecurity Engineer at Caixa Geral de Depósitos. He holds an MSc in Computer Engineering from the Faculty of Sciences of the University of Lisbon. He is a dedicated cybersecurity professional with a strong interest in securing complex systems and promoting cyber resilience.

\subsection*{  }
\setlength\intextsep{0pt} 
\begin{wrapfigure}{l}{0.13\textwidth}
    \centering
    \includegraphics[width=0.15\textwidth]{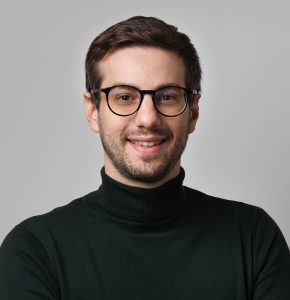}
\end{wrapfigure}
\noindent \textbf{Vinicius V. Cogo}
 is an Assistant Professor of the DI/CIÊNCIAS/ULisboa, has a PhD and a MSc in Informatics from CIÊNCIAS/ULisboa and a BSc in Computer Science from UFSM. He is also an integrated researcher at LASIGE, authored more than twenty peer-reviewed publications, and worked in many European projects in the areas of distributed systems, dependability, fault tolerance, storage of critical data, and cloud computing. More information at \url{https://www.di.fc.ul.pt/~vcogo/}.

\subsection*{  }
\setlength\intextsep{0pt} 
\begin{wrapfigure}{l}{0.13\textwidth}
    \centering
    \includegraphics[width=0.15\textwidth]{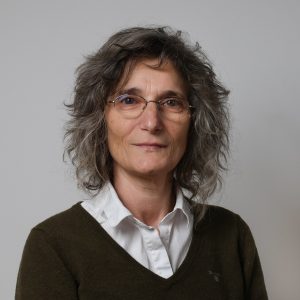}
\end{wrapfigure}
\noindent \textbf{Ib\'{e}ria Medeiros}
is an Associate Professor of the DI/CIÊNCIAS/ULisboa, and a member of the LASIGE research unit. Her research interests are software security, vulnerability and attack detection, code privacy and correction, and artificial intelligence applied to cybersecurity. She is the author of tools for software security and cybersecurity, and has been involved in international and national projects. More information about her at \url{https://www.di.fc.ul.pt/~imedeiros/}.

\end{document}